\documentclass[12pt, letterpaper]{article}

\usepackage{latexsym,multirow}
\usepackage{amssymb,amsmath, bm, bbm, dsfont, eufrak}
\usepackage{graphicx}
\usepackage{verbatim}
\usepackage{centernot}
\usepackage{multicol}
\usepackage{mathrsfs}
\usepackage[export]{adjustbox}
\usepackage{yfonts}

\usepackage{natbib}
\usepackage[pdftex, bookmarksopen=true, bookmarksnumbered=true,
pdfstartview=FitH, breaklinks=true, urlbordercolor={0 1 0}, citebordercolor={0 0 1}, hidelinks]{hyperref}
\usepackage{colortbl}
\usepackage{subfigure}
\usepackage{inputenc}
\usepackage{relsize}

\usepackage{accents}

\usepackage{dcolumn}
\newcolumntype{.}{D{.}{.}{-1}}
\newcolumntype{d}[1]{D{.}{.}{#1}}
\usepackage{theorem}
\theorembodyfont{\normalfont}
\theoremheaderfont{\normalfont\bfseries}
\newtheorem{theorem}{Theorem}

\newtheorem{assumption}{Assumption}
\theoremstyle{definition}
\newtheorem{definition}{Definition}

\newtheorem{lemma}{Lemma}
\newcommand{\qed}{\hfill \ensuremath{\Box}}

\usepackage{rotating}

\usepackage{arydshln}

\usepackage[compact]{titlesec}

\usepackage{array}
\newcolumntype{L}[1]{>{\raggedright\let\newline\\\arraybackslash\hspace{0pt}}m{#1}}
\newcolumntype{C}[1]{>{\centering\let\newline\\\arraybackslash\hspace{0pt}}m{#1}}
\newcolumntype{R}[1]{>{\raggedleft\let\newline\\\arraybackslash\hspace{0pt}}m{#1}}

\usepackage{booktabs}

\usepackage{tcolorbox}

\DeclareMathAlphabet{\mathpzc}{OT1}{pzc}{m}{it}

\allowdisplaybreaks 

\begin{document}

\newcommand\ud{\mathrm{d}}
\newcommand\dist{\buildrel\rm d\over\sim}
\newcommand\ind{\stackrel{\rm indep.}{\sim}}
\newcommand\iid{\stackrel{\rm i.i.d.}{\sim}}
\newcommand\logit{{\rm logit}}
\renewcommand\r{\right}
\renewcommand\l{\left}
\newcommand\pre{{(t-1)}}
\newcommand\cur{{(t)}}
\newcommand\cA{\mathcal{A}}
\newcommand\cB{\mathcal{B}}
\newcommand\bone{\mathbf{1}}
\newcommand\E{\mathbb{E}}
\newcommand\Var{{\rm Var}}
\newcommand\cD{\mathcal{D}}
\newcommand\cK{\mathcal{K}}
\newcommand\cP{\mathcal{P}}
\newcommand\cT{\mathcal{T}}
\newcommand\cX{\mathcal{X}}
\newcommand\cXR{\mathcal{X,R}}
\newcommand\wX{\widetilde{X}}
\newcommand\wT{\widetilde{T}}
\newcommand\wY{\widetilde{Y}}
\newcommand\wZ{\widetilde{Z}}
\newcommand\bX{\mathbf{X}}
\newcommand\bT{\mathbf{T}}
\newcommand\bt{\mathbf{t}}
\newcommand\bwT{\widetilde{\mathbf{T}}}
\newcommand\bwt{\tilde{\mathbf{t}}}
\newcommand\bbT{\overline{\mathbf{T}}}
\newcommand\bbt{\overline{\mathbf{t}}}
\newcommand\ubT{\underline{\mathbf{T}}}
\newcommand\ubt{\underline{\mathbf{t}}}
\newcommand\bhT{\widehat{\mathbf{T}}}
\newcommand\bht{\hat{\mathbf{t}}}

\newcommand\cF{\mathcal{F}} 
\newcommand\cC{\mathcal{C}} 
\newcommand\cS{\mathcal{S}} 
\newcommand\cN{\mathcal{N}} 
\newcommand\bZ{\mathbf{Z}} 
\newcommand\bz{\mathbf{z}} 

\newcommand\cw{\mathcal{w}} 
\newcommand\cW{\mathcal{W}} 
\newcommand\cZ{\mathcal{Z}} 

\newcommand\bw{\mathbf{w}} 
\newcommand\bW{\mathbf{W}} 

\newcommand\bg{\bar{g}} 
\newcommand\pg{g^\prime} 
\newcommand\brm{\bar{m}} 

\newcommand\cG{\mathcal{G}} 
\newcommand\cH{\mathcal{H}} 
\newcommand\cU{\mathcal{U}} 

\newcommand\pG{\mathpzc{G}} 
\newcommand\sG{G\leq g_s} 

\newcommand\pS{\mathscr{S}}
\newcommand\pP{\mathscr{P}}

\newcommand\cM{\mathcal{M}} 
\newcommand\cO{\mathcal{O}} 
\newcommand\cJ{\mathcal{J}} 

\newcommand\CDE{{\rm {\bf CDE}}}

\newcommand\cADE{{\rm {\bf cADE}}}

\newcommand\mo{\mathbf{1}}

\newcommand\gs{\{g^H, g^L\}}
\newcommand\ag{g^H}
\newcommand\abg{g^L}

\newcommand\au{u^H}
\newcommand\abu{u^L}

\newcommand\ADE{\delta}
\newcommand\DE{\delta}
\newcommand\NSE{\tau}
\newcommand\ANSE{\tau}
\newcommand\ATSE{\psi}

\newcommand\cANSE{{\rm {\bf cANSE}}}
\newcommand\ASE{{\rm {\bf ASE}}}
\newcommand\MSE{{\rm {\bf MSE}}}
\newcommand\INT{{\rm {\bf INT}}}

\newcommand\MDE{{\rm {\bf MDE}}}
\newcommand\CSE{{\rm {\bf CSE}}}
\newcommand\TSE{{\rm {\bf TSE}}}

\newcommand\Prz{{\rm Pr}_{\mathbf{z}}}

\newcommand\Pri{{\rm Pr}_i}

\newcommand\MR{{\rm MR}}
\newcommand\RR{{\rm RR}}

\newcommand\Nor{{\rm Normal}}

\newcommand\cATIE{{\sc cATIE}} 
\newcommand\ATIE{{\sc ATIE}}
\newcommand\cg{\cellcolor[gray]{0.7}}

\newcommand{\argmax}{\operatornamewithlimits{argmax}}
\newcommand{\argmin}{\operatornamewithlimits{argmin}}

\newcommand{\minb}{\operatornamewithlimits{min}}
\newcommand{\maxb}{\operatornamewithlimits{max}}

\newcommand\spacingset[1]{\renewcommand{\baselinestretch}%
{#1}\small\normalsize}

\setlength{\footnotesep}{\baselineskip}

\spacingset{1.25}

\newcommand{\tit}{{Spillover Effects in the Presence
    of Unobserved Networks}}



\title{\vspace{-0.6in}\bf \tit \thanks{I thank Alexander Coppock, Andrew Guess, and
    John Ternovski for providing me with data and
    answering my questions. I also thank Peter
    Aronow, Neal Beck, Forrest Crawford, Christian Fong,
    Erin Hartman, Zhichao Jiang, Marc Ratkovic, Cyrus Samii, Fredrik S{\"a}vje, Matthew Salganik, 
    members of Imai research group, and participants of the seminars at Yale
    University and New York University for
    helpful comments. I am particularly grateful to Kosuke Imai,
    Brandon Stewart, Santiago Olivella, and Adeline Lo for their
    continuous encouragement and detailed comments. 
    The earlier draft of this article was entitled ``Unbiased Estimation and Sensitivity Analysis for
    Network-Specific Spillover Effects: Application to An Online
    Network Experiment''\citep{egami2017unbiased}. The replication
    materials are available as \cite{egami2020}.} }

\author{Naoki Egami\thanks{Assistant Professor, Department of
      Political Science, Columbia University, New York NY 10027. \hspace{0.2in}
      Email:
      \href{mailto:naoki.egami@columbia.edu}{naoki.egami@columbia.edu}, URL:
      \url{https://naokiegami.com}}}

  \date{This Version: May 19, 2020 \\
    Forthcoming in {\it Political Analysis}}

\maketitle





\pdfbookmark[1]{Title Page}{Title Page}

\thispagestyle{empty} 
\setcounter{page}{0} 

\begin{abstract}
  When experimental subjects can interact with each other, the outcome of one
  individual may be affected by the treatment status of others. In many social science experiments, such spillover
  effects may occur through multiple networks, for example,
  through both online and offline face-to-face networks in a Twitter
  experiment. Thus, to understand how people use
  different networks, it is essential to estimate the spillover effect in each
  specific network separately. However, the unbiased estimation of these {\it network-specific spillover effects} requires 
  an often-violated assumption that researchers observe all relevant
  networks. We show that, unlike conventional omitted variable
  bias, bias due to unobserved networks remains 
  even when treatment assignment is randomized and when unobserved
  networks and a network of
  interest are independently generated. We then develop parametric and nonparametric sensitivity analysis methods,
  with which researchers can assess the potential influence of unobserved networks on
  causal findings. We illustrate the proposed methods with a
  simulation study based on a real-world Twitter network and an
  empirical application based on a network field experiment in China.
\end{abstract}
\noindent%
\vspace{0.1in}
\vfill

\newpage
\spacingset{1.55} 

\section{Introduction}
\label{sec:intro}
Existing methodologies for causal inference often assume the absence
of spillover effects, that is, people are affected only by treatments
directly assigned to them and not by those assigned to
others \citep{cox1958planning, rubin1980SUTVA}. However, in typical
social science experiments where individuals can interact with each
other, spillover effects might naturally arise, and this causal interdependence across people is often of theoretical
interest. Indeed, a growing number of political
science studies use experiments to estimate spillover effects on a
variety of outcomes, such as voting behavior
\citep{nickerson2008voting, sinclair2012social, sinclair2012detecting, foos2017all}, electoral
irregularities \citep{ichino2012deterring, asunka2017electoral, bowers2018models}, the
responsiveness of legislators \citep{coppock2014information},
information diffusion in ethnic networks \citep{larson2017ethnic}, and social norms in schools \citep{paluck2016changing}.

In many of such social science applications, people can interact with each
other through multiple channels, and thus, spillover effects often
arise through multiple networks. For example, even
though online network experiments typically focus only on online
networks, people can also communicate with each other
through their offline face-to-face network \citep{bond2012exp,
  aral2016networked, guess2016treatments, eckles2017Online}. Other
common types of multiple networks include friendship, neighbors,
ethnic, kinship, and partisan networks, among many others
\citep{fowler2011causality, sinclair2012social}. Therefore, it is important to estimate the spillover effect 
{\it specific} to each network. A substantive question is often not only about
whether there exists any spillover effect but also about which networks people use to share
information and influence each other's behavior. By estimating
{\it network-specific spillover effects}, researchers can examine the mechanism of spillover effects. 

However, the unbiased estimation of these network-specific spillover effects is challenging in
practice. It requires making an often-violated assumption that researchers observe all relevant
networks. For example, while scholars might carefully measure an ethnic network in a field experiment on ethnic voting, they might be
unable to measure other types of network interactions, such as neighbors, friendship, and kinship networks. In this
case, existing approaches, which assume no unobserved networks, can
misattribute the spillover effects in unobserved networks to the ethnic network, resulting in biased estimates of network-specific spillover effects.

In this article, we address this methodological challenge by formally
characterizing the bias due to unobserved networks and developing
sensitivity analysis methods, with which researchers can assess the
potential influence of unobserved networks on substantive findings. 

In Section~\ref{sec:multi}, we first extend the potential outcomes framework \citep{neyman1923,
  rubin1974causal} to settings with multiple networks and then formally define the {\it average
  network-specific spillover effect} (ANSE). This new estimand represents the average
causal effect of changing the treatment status of neighbors
in a given network, without changing the treatment status of neighbors
in other networks. The ANSE can be estimated without bias using an inverse probability
weighting estimator as long as researchers can observe the network of interest and all other networks in
which spillover effects exist. In Section~\ref{sec:bias}, we then consider
a scenario of unobserved networks and derive the exact bias
formula. It is a function of the spillover effects through unobserved networks and 
the overlap between observed and unobserved networks. This bias formula
implies that, unlike the conventional omitted variable bias, the bias for the ANSE is non-zero even
when (a) treatment assignment is randomized and (b) unobserved networks and the network of interest are independently generated. 

In Section~\ref{sec:sen}, we propose parametric and nonparametric 
sensitivity analysis methods for evaluating the potential influence of
unobserved networks on causal conclusions. Using these methods, researchers can derive 
simple formal conditions under which unobserved networks would 
explain away the estimated ANSE. Researchers can also use them
to bound the ANSE using only two sensitivity parameters. 
Although the parametric sensitivity analysis method focuses on one unobserved network for the
sake of intuitive interpretation, the nonparametric method can handle
multiple unobserved networks. 

We provide two empirical illustrations, each from the most popular types of network experiments; an online social network
experiment and a network field experiment. The first is a simulation
study based on the real-world Twitter network
\citep{guess2016treatments} where we simulate a variety of
unobserved offline face-to-face networks to assess the performance of
the proposed approach (Section~\ref{sec:sim}). The second is a reanalysis of the
field network experiment in rural China \citep{cai2015social}. We
apply the proposed sensitivity analysis methods and assess the
robustness of the original findings to unobserved networks (Section~\ref{sec:app}).

Our paper builds on a growing literature on spillover effects \citep[e.g.,][]{hong2006kinder, sobel2006, rosenbaum2007,
  hudgens2008toward, vaderweele2010interference}, especially spillover
effects in networks \citep[e.g.,][]{aronow2012random, bowers2013interference, toulis2013interference, hudgens2014,
  ogburn2014DAG, athey2016exact, forastiere2016observational,
  aronow2012interference, eckles2017design, tchetgen2017auto, bowers2018models}.
See \cite{halloran2016review} for a recent review about spillover
effects in general.  The vast majority of the work has mainly focused on the case where all relevant networks are observed.
Only recently has the literature begun to study the consequence of
unobserved networks. One way to handle unobserved networks is to
consider the problem as ``misspecification'' of the spillover
structure \citep{aronow2012interference, savje2019causal}. Although Proposition 8.1 of
\cite{aronow2012interference} implies that the inverse probability
weighting estimator is biased for the ANSE unless all relevant
networks are observed, the exact expression of bias is difficult to
characterize in general settings. In this paper, we instead focus on one common
type of misspecification -- the network of interest is observed, but
other relevant networks are unobserved. We, therefore, can explicitly
derive the exact bias formula and develop sensitivity analysis methods. 

Other approaches to deal with unobserved
networks include randomization tests \citep{luo2012inference, rosenbaum2007}
and the use of a monotonicity assumption
\citep{choi2016monotone}. While these approaches are robust to
unmeasured networks, estimands studied in these papers are designed to
detect the total amount of spillover effects in all networks, rather than
spillover effects specific to a particular network, which are the main
focus of this paper. \cite{savje2017average} discuss the estimation of
the expected average treatment effect in the presence of unknown
interference. While their result applies to general types of interference, their causal estimand is different
from ours, i.e., the direct effect of treatments rather than the
spillover effect. \cite{shpitser2019interference} propose to estimate underlying
networks by structural learning algorithms under the assumption of
a chain graph model. We take a different approach of the
potential outcomes framework, and our focus is to characterize the
exact bias and develop sensitivity analysis methods rather than
recovering underlying networks. 

\section{Spillover Effects in Multiple Networks}
\label{sec:multi}
Causal inference in randomized experiments often assumes no
interference \citep{cox1958planning, rubin1980SUTVA}, that is, units
are affected only by treatments directly assigned to them, and not by
treatments assigned to other units. However, when units are connected
in networks, treatments can have spillover effects on other units, and
this causal interdependence is of theoretical interest. While the
recent literature on interference has focused on settings with one network,
this section extends the potential outcomes framework to settings with multiple networks. This
setup serves as the foundation for analyzing unobserved networks and
developing methodologies in Sections~\ref{sec:bias} and~\ref{sec:sen}.

As our running example, we consider an online social network
experiment \citep{aral2016networked, eckles2017Online}, such as
political mobilization experiments on Facebook
\citep[e.g.,][]{bond2012exp} and on Twitter
\citep[e.g.,][]{guess2016treatments}. Although typical online network
experiments only measure online networks of interest, people are often
embedded in other networks, most importantly, an offline face-to-face network. Thus, experimental subjects can share
information with one another through Facebook as well as through their
offline face-to-face interactions, inducing spillover effects in both
online and offline networks. To introduce the multiple
network framework, we focus on such spillover effects in online and offline face-to-face networks as an illustrative example. In
Sections~\ref{sec:sim} and~\ref{sec:app}, we provide 
two detailed empirical illustrations; a simulation study based on the Twitter
network \citep{guess2016treatments} 
and a reanalysis of the field network experiment in rural China
\citep{cai2015social}, respectively.

\subsection{The Setup}
\label{subsec:notation}
Consider a randomized experiment with sample size $N$ and each unit is
indexed by $i \in  \{1,2, \ldots, N\}$. 
Define a treatment assignment vector $\bT = (T_1, \ldots, T_N)^\top$ where binary treatment variable $T_i \in \{0,
1\}$ denotes the treatment received by unit $i$. For example, $T_i =
1$ would mean that unit $i$ receives an informational message, 
and $T_i = 0$ would mean unit $i$ is in a control group who does not receive any message. Based on the experimental
design, treatment assignment probability $\Pr(\bT=\bt)$ is known
for all $\bt \in \{0,1\}^N$. Using the potential outcomes framework \citep{neyman1923,
  rubin1974causal}, let $Y_i (\bt)$ denote the potential outcome of individual $i$ if the
treatment assignment vector $\bT$ is set to $\bt$.  Importantly, the potential 
outcome of individual $i$ is affected not only by her own treatment
assignment but also by the treatments received by others. Thus, we
allow for spillover/interference between individuals
\citep{cox1958planning, rubin1980SUTVA}. In our running example,
voting behavior of a given individual can be affected not only by
whether she directly receives the informational message but also by
messages assigned to her friends as people can share information with one another. 


\begin{figure}[!t]
  \begin{center}
    \hspace{0.5in}\includegraphics[width = 3in]{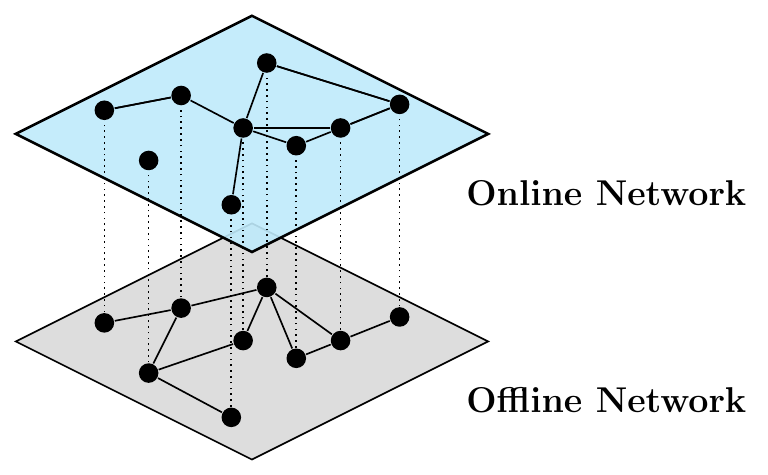}
  \end{center}  \vspace{-0.25in}    
  \spacingset{1}{
    \caption{Example of Two Networks. {\it Note}: The same set of
      people are connected differently by an online network and an offline network. Neighbors in the two networks overlap, i.e.,
      some, but not all, pairs of people are connected in both
      online and offline networks.}\label{fig:net}
    }
\end{figure}

To formalize whose treatment status can affect a given individual, we rely on 
networks. In particular,  consider two networks $\cG$ and $\cU$
connecting units with different edge sets. For example, people can be
connected with Facebook and their offline face-to-face network. Formally, we define an adjacency  
matrix of network $\cG$ to be $A^{\cG}$ where $A^{\cG}_{ij} =
A^{\cG}_{ji}  = 1$ if unit $i$ is connected to unit $j$ (and zeros on the
diagonal). We can also incorporate directed ties when we define
$A^{\cG}_{ij} = 1$ if unit $i$ ``follows'' unit $j$, for example, on Twitter. By assigning different weights to each tie, this
framework can accommodate the strength of ties as well. We then define individual $i$'s {\it neighbors} in network
$\cG$ to be other individuals to whom she is connected in network $\cG$,
formally, $\cN^{\cG}_i \equiv \{j: A^{\cG}_{ij} \neq 0\}.$ We define
$A^{\cU}$ and $\cN^{\cU}_i$ similarly. For example, neighbors are
those with whom 
individual $i$ is friends on Facebook ($\cN^{\cG}_i$; neighbors on an online network) or those who individual $i$ meets in person on a daily
basis ($\cN^{\cU}_i$; neighbors on an offline face-to-face network). Importantly, neighbors in the two networks can overlap; we
meet some but not all of our Facebook friends on a daily
basis. Figure~\ref{fig:net} visualizes an example with two networks.

We explicitly incorporate these two networks into the potential
outcomes. In particular, we extend the stratified
interference assumption \citep{hudgens2008toward} to multiple
networks. We assume that the potential outcomes of
individual $i$ are affected by her own treatment assignment and the
treated proportions of neighbors in networks $\cG$ and $\cU$ 
\citep{manski2013identification, toulis2013interference, hudgens2014, forastiere2016observational}. 
\begin{assumption}[Stratified Interference with Multiple Networks]
  \label{mean} $ $ \\ For all $i,$ 
  \begin{equation*} 
    Y_i (\bt) = Y_i (d, g, u),
  \end{equation*}
  where $d$ is her own treatment assignment, $g$ and $u$ are the
  proportions of treated neighbors in networks $\cG$ and $\cU$,
  respectively:
  \begin{align*}
    d &  = t_i, \ \  g  = \frac{\sum_{j \in \cN_i^{\cG}}
        t_j}{|\cN_i^{\cG}|}, \ \ u  = \frac{\sum_{j \in \cN_i^{\cU}} t_j}{|\cN_i^{\cU}|},
\end{align*}
where $t_i$ is the treatment received by individual $i$. $\sum_{j \in
  \cN_i^{\cG}} t_j$ is the total number of treated units among individual
$i$'s neighbors in network $\cG$ and $|\cN_i^{\cG}|$ is the total
number of individual $i$'s neighbors in network $\cG$. $\sum_{j \in
  \cN_i^{\cU}} t_j$ and $|\cN_i^{\cU}|$ are similarly defined.  
\end{assumption}
The potential outcomes depend on the treatment assignment of
herself $d$ and the fractions of treated neighbors in each network
$(g,u)$. For example, the potential outcome of unit $i$ is a function
of whether she receives the informational message ($d$), the
proportion of Facebook friends who receive the message ($g$), and the
proportion of offline friends who receive the message ($u$).  
Although this assumption can only allow for spillover effects
through treated proportions in multiple networks, it is 
more flexible than the conventional assumption of no interference,
which requires that the treatment status of neighbors do not change
potential outcomes at all. Formally, Assumption~\ref{mean} can also be viewed as
an exposure mapping $f(\bt, (A^\cG_{i}, A^\cU_{i}))$
set to a three-dimensional vector $(d, g, u)$
\citep{aronow2012interference}. 

Several points about Assumption~\ref{mean} are worth clarifying. First, because this
assumption is made at the individual level, an equivalent statement of
Assumption~\ref{mean} is that the potential outcomes depend on 
the treatment assignment of herself and the number (rather than
proportions) of treated neighbors in each network. The only change is
that when defining causal estimands, we should explicitly condition on
the total number of neighbors in each network. Second,
Assumption~\ref{mean} is violated when neighbors in the same network have different
spillover effects. For example, suppose a given individual $i$ has
three Facebook friends $\{F_1, F_2, F_3\}$, and $F_1$ has a larger
spillover effect than the other two. In this case, even when $g = 1/3$,
the potential outcomes of unit $i$ differ depending on whether $F_1$ is treated or
one of the other two is treated, which violates
Assumption~\ref{mean}. Third and most importantly, unlike the previous
literature, Assumption~\ref{mean} is defined with multiple
networks. Therefore, researchers can make the assumption more
plausible by specifying networks more precisely. In the example above, suppose $F_1$ has a
larger spillover effect than the other two because $F_1$ is also a
friend of unit $i$ on Twitter but the other two are not. Then, we can restore
Assumption~\ref{mean} with respect to three networks; the Facebook
network, the Twitter network, and an offline network. However, it is
important to note that as we consider more networks, it is more
difficult to observe all relevant networks, which is the central topic
of the paper we further discuss in Sections~\ref{sec:bias} and~\ref{sec:sen}.

We define networks to be {\it causally relevant} if treated
proportions in such networks have non-zero causal effects. \vspace{-0.1in}
\begin{definition}[Relevant Networks] \label{rel}
  Network $\cG$ is causally relevant if $\ Y_i (d, g, u) \ \neq \ Y_i (d,
  g^\prime, u) \ $ for some $i$ and $(d, g, u), (d, g^\prime, u) \in
  \Delta_i$ where $\Delta_i$ is the support of treatment exposure
  vector $(T_i, G_i, U_i)$. The relevance of network $\cU$ is similarly defined. 
\end{definition}
For example, if people do not talk about elections at all on Facebook,
Facebook would be causally irrelevant because messages received by
other Facebook friends would have no effect on voting behavior through
Facebook when fixing treated proportions in the offline network.  

Finally, as in the standard causal inference setting, we observe only one
of many potential outcomes. For all $i$, $ Y_i = \sum_{(d,g,u) \in \Delta_i} \mo\{T_i=d, G_i=g,
U_i=u\} \ Y_i(d,g,u)$ \citep{holland1986statistics}. 

\subsection{Causal Quantities of Interest}
\label{subsec:ADEANSE}
Without loss of generality, we assume two networks $\cG$ and $\cU$ are
causally relevant and define our quantities of interest using these two networks. 

\subsubsection{Direct Effects and Network-Specific Spillover Effects}
\label{subsubsec:ANSE}
First, we define the direct effect of a treatment. It is
the difference between the potential outcomes under treatment and control, averaging over the distribution
of treatment assignment $(G_i, U_i)$. Formally, we define the {\it
  average direct effect} (ADE) as follows.
\begin{definition}[Average Direct Effect]
  \begin{equation}
    \DE \ \equiv \ \frac{1}{N} \sum_{i=1}^N \l\{\sum_{(g,u) \in \Delta_i^{gu}} \{Y_i(1,  g, u) -
    Y_i(0, g, u)\} \Pr(G_i=g, U_i=u)\r\},
  \end{equation}
  where the support is defined  as $\Delta_i^{gu} = \{(g,u):\Pr(G_i=g, U_i=u) > 0\}.$ 
\end{definition}
It is the weighted average of the 
causal effects, $Y_i(1, g, u) - Y_i(0, g,u)$, which hold the
proportions of treated neighbors in the two networks constant. 
Intuitively, this effect quantifies the causal impact of the treatment
directly received by herself. Thus, the ADE could represent the direct
causal effect of the informational message on voting behavior. The ADE is a simple extension of the expected average
treatment effect \citep{savje2017average} to the two-network
settings. We keep the term of the {\it direct} effect in order to distinguish it from spillover effects we define next. 

Spillover effects describe the causal effects of the neighbors' treatment status on a given 
individual. It is the change in the potential outcome when
the proportion of treated neighbors goes from a lower proportion  
to a higher proportion. Formally, it is the difference between the potential outcome for
a given individual when $\ag \times 100$ percent of her neighbors in $\cG$ are
treated and the other potential outcome for the same individual when
$\abg \times 100$ percent of her neighbors in $\cG$ are
treated, holding her own treatment assignment and averaging over the proportions of treated
neighbors in $\cU$. Two constants $g^H$ and $g^L$ stand for ``Higher'' and
``Lower'' percent. We define the {\it
  average network-specific effect} (ANSE) as follows.
\begin{definition}[Average Network-Specific Spillover Effect] \label{def:ANSE}
\begin{equation}
  \NSE (\ag, \abg ; d)  \ \equiv \ \frac{1}{N} \sum_{i=1}^N \l\{\sum_{u \in \Delta_i^u} \{Y_i(d, \ag, u) -
  Y_i(d,  \abg, u) \} \ \Pr (U_i=u \mid T_i=d, G_i = g^L)\r\}, \label{eq:unitANSE}
\end{equation}
where the support is defined as $\Delta_i^{u} = \{u: \Pr(U_i=u \mid
T_i=d, G_i = g^H) > 0 \mbox{ and }  \Pr(U_i=u \mid T_i=d, G_i = g^L) > 0\}.$ The ANSE for network $\cU$ is
defined similarly. 
\end{definition}
It is the weighted average of the spillover
effects specific to network $\cG$, $Y_i(d, g^H, u) - Y_i(d, g^L,u)$,
which hold her own treatment assignment and the proportion of treated
neighbors in $\cU$ constant. Thus, for example, the ANSE $(\ag=0.8, \abg=0.2; d = 0)$ could represent the
Facebook-specific spillover effect where we change the treated proportion
of one's Facebook friends (from $g^L = 20$\% to $g^H = 80$\%) while holding constant
one's own treatment assignment ($d = 0$) and averaging over the treated proportion of
one's offline friends ($u$). Similarly, the ANSE  $(u^H =0.8, u^L=0.2;
d = 0)$ could represent the offline network-specific spillover effect where we change the treated proportion
of one's offline friends (from $u^L = 20$\% to $u^H = 80$\%) while holding constant
one's own treatment assignment ($d = 0$) and averaging over the treated proportion of
one's Facebook friends ($u$). 

Importantly, the ANSE captures the spillover effect specific to each network separately. Thus, when
Facebook (the offline network) is causally irrelevant, as defined in Definition~\ref{rel}, the ANSE in Facebook
(the offline network) will be zero. By estimating the ANSE for each network, researchers can unpack the mechanism
of spillover effects; which networks do people use to share
information and influence each other's voting behavior? 
In Appendix~\ref{sec:decom}, we additionally  show that the ANSE decomposes the total
spillover effect, a popular estimand in the literature
\citep{hudgens2008toward}, into each network. 

\subsection{Estimation}
\label{subsec:estimation}
In this section, we study the estimation of the ADE and the ANSE. We
begin by introducing a common assumption made in practice; the {\it no
  omitted network} assumption defined below.
\begin{assumption}[No Omitted Network]
  \label{noOmitted} All causally relevant networks are observed.
\end{assumption}
For example, in the Facebook mobilization experiment \citep{bond2012exp}, 
this assumption of no omitted network means that the Facebook network is the only relevant network
and an unobserved face-to-face network is irrelevant; experimental
subjects could affect one another through Facebook but not through
their unobserved offline interactions. We first examine the estimation
under this assumption and then discuss its violation in the subsequent
sections.

Following the recent literature \citep[e.g.,][]{hudgens2008toward,
  vaderweele2010interference, kang2016peer, 
  airoldi2017elements}, we start with
design-based, inverse probability weighting (IPW) estimators for the
ADE and ANSE \citep{horvitz1952generalization, aronow2012interference}. Importantly,
researchers can compute the treatment exposure probability $\Pr(T_i =
d, G_i = g, U_i = u)$ from the experimental design under the no
omitted network assumption.\footnote{It is important to emphasize that
  the treatment exposure probability is a function of both network
  structure and experimental design.} Thus, we rely on the following weighting estimators. 

\begin{theorem}[Estimation of the ADE and the ANSE]
  \label{estimation}
  Under Assumption~\ref{noOmitted}, the treatment exposure
  probability $\Pr(T_i=d, G_i=g, U_i=u)$ is known for all $(d, g, u)
  \in \Delta_i$ and all $i$. Therefore, the average
  direct effect and the average network-specific spillover effect in $\cG$
  can be estimated by the following inverse probability weighting
  estimators. 
  \begin{eqnarray*}
    \E[\widehat{\ADE}]   =  \ADE \ \ &\mbox{ and } &  \ \  \E[\widehat{\ANSE}(\ag, \abg; d) ]   =  \ANSE(\ag, \abg; d),
  \end{eqnarray*}
    where the expectation is taken over the experimental design
    $\Pr(\bT=\bt)$, and 
  \begin{eqnarray}
    \widehat{\ADE} &\equiv& \frac{1}{N} \sum_{i=1}^N \mo\{T_i=1\}
                            \widetilde{w}_i Y_i - \frac{1}{N}
                            \sum_{i=1}^N \mo\{T_i=0\} \widetilde{w}_i
                            Y_i \\
    \widehat{\ANSE}(\ag, \abg; d)  & \equiv & \frac{1}{N} \sum_{i=1}^N \mo\{T_i = d, G_i = \ag\}
                                w_i Y_i - \frac{1}{N} \sum_{i=1}^N \mo\{T_i = d, G_i = \abg\} w_i Y_i, 
  \end{eqnarray}
  and weights are defined as
  \begin{eqnarray*}
    \widetilde{w}_i = \frac{1}{\Pr(T_i \mid G_i, U_i)}, 
   \ \ &\mbox{ and } &  \ \ 
    w_i = \frac{\Pr(U_i \mid T_i, G_i  = \abg)}{\Pr(U_i \mid T_i, G_i)} \times \frac{1}{\Pr(T_i, G_i)}.
  \end{eqnarray*}
\end{theorem}
We provide the proof in Appendix~\ref{app:identification}. Although the IPW estimators are unbiased, researchers might want to
increase efficiency by incorporating covariate information and some
parametric assumptions \citep{sarndal1992model}. In particular, 
we build on the design-based IPW estimator above and propose a weighted
linear regression estimator. This alternative approach reduces
standard errors at the expense of some bias due to parametric modeling
assumptions. In particular, we consider 
\begin{equation}
  Y_i \sim \alpha + \beta T_i + \gamma_0 G_i + \gamma_1 (T_i \times G_i) + \bX^\top_
  i \theta \mbox{    with weights  } w_i. 
\end{equation}
The key assumptions are the linearity of $G_i$ and the inclusion of pre-treatment
covariates $\bX_i$.\footnote{It is also possible to incorporate
  higher-order polynomial terms of $G_i$.} We emphasize an important
tradeoff. On the one hand, both assumptions are parametric in the sense that they are not
directly justified by the experimental design. On the other hand, they tend to
greatly reduce variance. Especially in settings with multiple
networks, treatment exposure probabilities $\Pr(T_i, G_i, U_i)$ are
usually small, and in many applied contexts, the standard IPW estimator
has large standard errors. The weighted linear
regression estimator aims to balance this bias-variance tradeoff in
practice. See \cite{toulis2013interference} for a Bayesian approach,
and \cite{sarndal1992model, rosenbaum2002covariance} and
\cite{aronow2012interference} for the design-based covariate
adjustment.  

\section{Exact Bias Formula}
\label{sec:bias}
We can obtain an unbiased estimate of the ANSE when we observe all relevant
networks. However, this assumption of no omitted network is often
violated in practice, and if so, the ANSE even 
in the observed network cannot be estimated without bias. For example,
in \cite{bond2012exp}, while the Facebook network is observed, an
offline face-to-face network is unobserved. In this case, even with
randomized experiments, an estimate of the Facebook-specific spillover
effect is biased because people can potentially share information with
offline friends. \cite{bond2012exp} write, ``it is plausible that unobserved face-to-face interactions account for at least some of the
social influence that we observed in this experiment'' (p. 298). In
this section, to explicitly characterize sources of such bias, we derive
the exact bias formula for the ANSE. We provide the exact bias formula
for the ADE in Appendix~\ref{app:biasADE}. 

We consider a common research setting in which the main network of interest is observed but other relevant
networks are not observed.  In particular, we assume $\cG$ is an observed
network of interest and $\cU$ is unobserved.\footnote{If even the
main network of interest is partially unobserved, it is impossible to
identify the ANSE without strong assumptions because the treatment variable itself (the
proportion of treated neighbors in the main network) is unobserved. If
researchers are interested in estimating the ANSE in the observed part of the main
network, the same results in Sections~\ref{sec:bias}
and~\ref{sec:sen} hold by viewing the unobserved part of the main
network as a separate unobserved network.} Thus, the quantity of interest is the ANSE
in the observed network $\cG$.  For simplicity, we refer to the ANSE in
$\cG$ as the ANSE, without explicitly mentioning $\cG$.

Since network $\cU$ is unmeasured, we cannot use weights $w_i$
discussed in Section~\ref{subsec:estimation}. Instead, we can only
rely on partial weights $ w^{\texttt{B}}_i \equiv 1/\Pr(T_i, G_i)$ 
where we only adjust for the direct treatment assignment $T_i$ and the
treated proportion in the observed network $G_i$. For example, the IPW
estimator based on such partial weights becomes 
\begin{equation}
  \frac{1}{N} \sum_{i=1}^N 
  \biggl\{ \mo\{T_i=d, G_i=\ag\}w_i^{\texttt{B}} Y_i - \mo\{T_i=d, G_i=\abg\}w_i^{\texttt{B}} Y_i\biggr\}.\label{eq:biasANSE}
\end{equation}

We use $\widehat{\ANSE}_B (\ag, \abg; d)$ to denote any estimator with
partial weights $w^{\texttt{B}}_i$, including both the IPW estimator and the
regression estimator discussed in Section~\ref{subsec:estimation}. For the regression estimator, the bias
formula we derive next can be seen as the lower bound where we
focus only on the bias due to unobserved networks and not on bias due
to functional form assumptions. 

The next theorem shows the exact bias formula for $\hat{\ANSE}_B (\ag,
\abg; d)$ in settings where the no omitted network assumption does not hold. 

\begin{theorem}[Bias for the ANSE due to Omitted Networks]
\label{biasASE}
When the no omitted network assumption (Assumption~\ref{noOmitted}) is
 violated, estimator $\hat{\ANSE}_B (\ag, \abg; d)$ is biased for the ANSE. 
\begin{eqnarray*}
  \hspace{-0.4in} && \E[\hat{\ANSE}_B (\ag, \abg; d)]  - \ANSE (\ag, \abg; d)\\[5pt]
  \hspace{-0.4in} &= & \hspace{-0.1in}\frac{1}{N} \sum_{i=1}^N \biggl\{ \sum_{u \in \Delta_i^u} \{Y_i (d, \ag, u) -
                       Y_i(d, \ag, u^\prime)\} \{ \Pr(U_i=u|T_i=d, G_i=\ag) -
                       \Pr(U_i=u|T_i=d, G_i = \abg)\}\biggr\},
\end{eqnarray*}
for any $u^\prime \in \Delta^u$.
\end{theorem}
We provide the proof in Appendix~\ref{app:biasASE}. This bias can be decomposed into two parts: (1) the spillover effects
attributable to the unobserved network $\cU$, $Y_i (d, \ag, u) - Y_i(d,
\ag, u^\prime)$; 
(2) the dependence between the fraction of treated neighbors in $\cG$
and the fraction of treated neighbors in $\cU$, $\Pr(U_i=u \mid T_i=d, G_i=\ag) -
\Pr(U_i=u \mid T_i=d, G_i = \abg)$. 

Based on this decomposition, we offer several implications of the theorem. First,
when treatment assignment to $\cU$ has no effect (i.e., $\cU$ is
irrelevant), $Y_i (d, \ag, u) -  Y_i(d, \ag, u^\prime) \ = \
0$ for any $i$. In this simple case, the bias is zero; this formula
includes the assumption of no omitted network as a special case. 

Second, the dependence between the fraction of treated neighbors in $\cG$
and the fraction of treated neighbors in $\cU$ determines the size and
sign of the bias. In theory, when $G_i$ and $U_i$ are independent given $T_i$, the bias is
zero. However, $G_i$ and $U_i$ are in general dependent given
$T_i$. Two points about this dependence are worth noting. First, some randomization of treatment assignment, such
as a Bernoulli design, can make $T_i$ independent of $(G_i, U_i)$, 
but $G_i$ and $U_i$ are dependent given $T_i$ even after any randomization of
treatment assignment because some neighbors in $\cG$ are also neighbors in
the other network $\cU$; networks $\cG$ and $\cU$ overlap each other. 
Formally, $G_i$ and $U_i$ are dependent because both
are functions of the treatment assignment to common neighbors. Second, based on the
same logic, $G_i$ and $U_i$ are not independent
given $T_i$ even when two networks $\cG$ and $\cU$ are
independently generated because the two networks can still overlap each other.

For example, the Facebook network and an 
unobserved face-to-face network are likely to overlap each
other. For some users, their Facebook friends are also close
friends to whom they have offline interactions and vice versa. As
long as the spillover effects in the face-to-face network are 
non-zero, an estimator ignoring this unobserved offline network
(equation~\eqref{eq:biasANSE}) would be
biased for the Facebook-specific spillover effect. 

Finally, we illustrate the bias with a simple linear model,  
$ Y_i (T_i, G_i, U_i) \ = \alpha + \beta T_i + \gamma G_i + \lambda U_i
+ \epsilon_i$ where $\epsilon_i$ is an error term. Under the model, the potential outcome of individual $i$ depends on neighbors
in both $\cG$ and $\cU$, and such two network-specific spillover
effects do not interact. In this simple setup, the bias can be simplified as follows. 
\begin{eqnarray*}
  \E[\hat{\ANSE}_B(\ag, \abg; d)] - \ANSE(\ag,\abg; d)
  &= &  \lambda \times \frac{1}{N} \sum_{i=1}^N \{\E[U_i \mid T_i=d, G_i=\ag] - \E[U_i
       \mid T_i=d, G_i=\abg]\}.
\end{eqnarray*}
It is clear that the bias depends on the ANSE in the
unobserved network $\cU$ (i.e., $\lambda$) and the association between $U_i$ and
$G_i$ given $T_i$. This bias is not zero unless the unobserved network
$\cU$ is irrelevant because $\E[U_i \mid  T_i=d, G_i=\ag] \ \neq \
\E[U_i \mid  T_i=d, G_i=\abg]$ in general.

\section{Sensitivity Analysis}
\label{sec:sen}
We address the potential violation of the no omitted network assumption by developing 
parametric and nonparametric sensitivity analysis methods for the ANSE.

\subsection{Parametric Sensitivity Analysis}
\label{subsec:parasen}
Although the exact bias formula in Theorem~\ref{biasASE} does
not make any assumption about the unobserved network $\cU$, in order to use
it in applied work, it requires specifying a large number of
sensitivity parameters. To construct a practical parametric sensitivity
analysis method, we rely on a simplifying parametric assumption.  In
particular, we consider the assumption that the network-specific spillover effect in an unobserved
network is linear and additive \citep{airoldi2017elements}.
\begin{assumption}[Linear, Additive Network-Specific Spillover Effect in $\cU$] 
  \label{linear} 
  \begin{equation*}
    \frac{1}{\sum_{i=1}^N \mo\{S_i=s\}}\sum_{i:S_i=s} \{Y_i(d, g, u) -
    Y_i(d, g, u^\prime)\} = \lambda (u- u^\prime), 
  \end{equation*}   
\end{assumption}
with some coefficient $\lambda$ for all $(d, g, u), (d, g, u^\prime) \in \Delta_s$ where
$\Delta_s$ is the support of $(d, g, u)$ for $i$ with the neighbors'
profile $S_i=s$. The neighbors' profile $S$ is defined such that the probability of
treatment exposure is the same for those who have the same neighbors'
profile. Formally, for $i \neq j$, $\Pr(T_i=d, G_i=g, U_i=u) = \Pr(T_j=d, G_j=g, U_j=u)$ when
$S_i=S_j.$ For example, when the Bernoulli or completely
randomized design is used, $S_i$ is simply a vector of three values; the number of neighbors
in $\cG$ and $\cU$, and the number of neighbors common to the two
networks.

While this linear additive assumption is strong, it is commonly used
in the causal inference literature to derive intuitive, easy-to-use
sensitivity analysis methods. For example, the widely-used sensitivity analysis
methods for unobserved confounders in observational studies rely on
similar assumptions \citep{vanderweele2011bias}. We also consider
nonparametric sensitivity analysis in the next section under weaker assumptions.

Under Assumption~\ref{linear}, the general bias formula
becomes the multiplication of two terms: the network-specific spillover
effect in $\cU$, i.e., $\lambda$, and the effect of $G_i$ on $U_i$
given $T_i$, i.e., $ \frac{1}{N} \sum_{i=1}^N \{ \E[U_i \mid T_i=d, G_i=\ag] - \E[U_i
\mid T_i=d, G_i=\abg] \}.$ The following theorem shows a simplified
bias formula for settings with sparse $\cG$.

\begin{theorem}[Parametric Sensitivity Analysis]
  \label{Parasen1}
  Under Assumption~\ref{linear}, a bias formula is approximated as follows. 
  \begin{equation}
    \E[\hat{\ANSE}_B (\ag, \abg; d)]  - \ANSE (\ag, \abg; d) \approx \lambda \times \pi_{GU} \times (\ag - \abg),
    \label{eq:biassen1}
  \end{equation}
  where $\lambda$ is the ANSE in network $\cU$ and 
  $\pi_{GU}$ is the overlap, i.e., the fraction of neighbors in $\cU$ who are also
  neighbors in $\cG$. Formally $\pi_{GU} \equiv \sum_{i=1}^N
  \{|\cN^{(\cG,\cU)}_i| / |\cN^{\cU}_i|\}/N$  where
  $|\cN^{(\cG,\cU)}_i|$ is the number of unit $i$'s neighbors common to the two
  networks. If an experiment uses a Bernoulli design, the
  approximation is exact regardless of the sparsity of network $\cG.$
\end{theorem}
We provide the proof in Appendix~\ref{app:sen1}. This simplified bias formula offers several implications. First,
the bias is small when $\pi_{GU}$ is small, i.e., the overlap of
neighbors in $\cG$ and $\cU$ is small. Hence, the bias is close to
zero when the network $\cG$ is sparse and neighbors in $\cG$ and $\cU$
are disjoint.  Second, even if two networks $\cG$ and $\cU$ are independently generated, the bias is not zero because
$\pi_{GU} \neq 0$. We derive a similar parametric bias formula for
settings with non-sparse $\cG$ in Appendix~\ref{app:sen1}. 

For example, the overlap between the Facebook network and the
unobserved face-to-face network is defined to be the fraction of  
offline friends who are also friends on Facebook. When this overlap is
large (small), we expect the bias for the Facebook-specific spillover effect to be large (small). 

To use this formula for a sensitivity analysis, researchers need to specify two
sensitivity parameters: the network-specific spillover effect in an unobserved network
(i.e., $\lambda$) and the fraction of neighbors in $\cU$ who are also
neighbors in $\cG $ (i.e., $\pi_{GU}$). Once these two parameters are
specified, we can derive the bias. Subsequently, because the bias
utilizes only sensitivity parameters and $(\ag - \abg)$, we can
obtain a bias corrected estimate by subtracting this bias
from $\hat{\ANSE}_B (\ag, \abg; d)$. A sensitivity analysis is
to report the estimated ANSE under a
range of plausible values of $\lambda$ and $\pi_{GU}$ where $0 <
\pi_{GU} < 1.$ Note that $\lambda = 0 $ corresponds to the no omitted
network assumption. 

\subsection{Nonparametric Sensitivity Analysis}
\label{subsec:nonparasen}
Now, we provide a nonparametric sensitivity analysis
that makes only one assumption of non-negative outcomes. 
While we introduce our method using a random variable $U_i$
for simplicity, the same method can be applied to a random vector
${\bf U_i}$ and accommodate multiple unobserved networks. The proposed
method is an extension of a sensitivity analysis developed for
observational studies with no spillover effect
\citep{ding2016sensitivity} to experimental settings where spillover effects in unobserved networks induce bias.

As the parametric sensitivity analysis, we use two
sensitivity parameters: intuitively, the network-specific spillover
effect in $\cU$, and the association between $U_i$ and
$G_i$. To capture the network-specific spillover effect in $\cU$, let $\MR_{UY} (g, s)\equiv
\maxb_{u} \sum_{i; S_i=s} Y_i(d, g, u) / \minb_{u}\sum_{i;S_i=s} Y_i(d,g,u)$ denote the
largest potential outcomes ratios of $U_i$ on $Y_i$ given $T_i=d,
G_i=g$ and $S_i=s$. For
notational simplicity, we drop subscript $d$ whenever it is obvious
from contexts. Then, we define $\MR_{UY} = \maxb_{g \in (\ag, \abg), \ s \in \cS}  \ \MR_{UY} (g, s)$ as the
largest potential outcomes ratio of $U_i$ on $Y_i$ over $g \in \gs$ and
$s \in \cS.$ Thus,
$\MR_{UY}$ quantifies the largest possible potential outcomes ratio of
$U_i$ on $Y_i$. This ratio captures the magnitude of the
network-specific spillover effect in $\cU$. When network $\cU$ is
irrelevant, $\MR_{UY} = 1$.

Furthermore, to quantify the association between $G_i$ and $U_i$, 
we use $\RR_{GU} (g, g^\prime ,u, s) \ = \ 
\Pr(U_i=u \mid  T_i=d, G_i=g) \ / \ \Pr(U_i=u \mid T_i=d, G_i=g^\prime)$ to
denote the relative risks of $G_i$ on $U_i$ for all $i$ with $S_i=s$.
$ \RR_{GU} \ = \ \maxb_{(g, g^\prime) \in (\ag, \abg), u \in \Delta^u, s
  \in \cS} \ \RR_{GU} (g, g^\prime ,u, s)$ is the
maximum of these relative risks. This risk ratio captures the
association between $G_i$ and $U_i$.

Using these ratios, we can derive an inequality that the ANSE in
the observed network $\cG$ needs to satisfy as long as outcomes are
non-negative. The next theorem shows that we can obtain the bound for
the ANSE with two sensitivity parameters, $\MR_{UY}$ and $\RR_{GU}$.

\begin{theorem}[Bound on the ANSE]
\label{sen2}
  When outcomes are non-negative, 
  \begin{eqnarray*}
    && \E\biggl[ \frac{\widehat{m}(d, \ag)}{B} -  B \widehat{m}(d,
       \abg) \biggr]
       \ \leq \  \ANSE (\ag, \abg; d) \ \leq \  \E\biggl[  B
       \widehat{m}(d, \ag) -  \frac{\widehat{m}(d, \abg)}{B} \biggr],
  \end{eqnarray*}
  where  $B = (\RR_{GU} \times \MR_{UY})/ (\RR_{GU} + \MR_{UY} - 1)$
  and $\widehat{m}(d, g) = \frac{1}{N} \sum_{i=1}^N \mo\{T_i=d,
    G_i=g\} w_i^{\texttt{B}}Y_i$ for $g \in \gs$.
\end{theorem}
We provide the proof in Appendix~\ref{app:sen2}. Note that $\MR_{UY} = 1 $ corresponds to the no omitted network
assumption, and $\E[\widehat{m}(d, \ag) - \widehat{m}(d, \abg)] =
\E[\hat{\ANSE}_B (\ag, \abg; d)] = \ANSE (\ag, \abg; d)$ under the
assumption. $B$ is an increasing function of both $\RR_{GU}$ and
$\MR_{UY}$, implying that the bound is wider when the
network-specific spillover effect in the unobserved network 
$\cU$ is larger and the effect of $G_i$ on the distribution of $U_i$ is larger. 
In fact, the size of the bound is given by $(B - \frac{1}{B})
\E[\widehat{m}(d, g^H) + \widehat{m}(d, g^L)].$ It is important to
note that the bound is not location-invariant because we use mean ratios
and risk ratios as sensitivity parameters
\citep{ding2016sensitivity}. The bound is valid as far as outcomes are
non-negative, but how informative it is can vary. 
To conduct the sensitivity analysis, one can compute the bound for a range of
plausible values of $\MR_{UY}$ and $\RR_{GU}$. Compared to
the parametric sensitivity analysis, $\MR_{UY}$ and $\RR_{GU}$
correspond to $\lambda$ and $\pi_{GU},$ respectively. Finally, as in
Section~\ref{subsec:estimation}, we can also use a weighted linear
regression estimator for $\widehat{m}(d, g)$ instead of the IPW estimator.

\section{Simulation Study: Twitter Mobilization Experiment}
\label{sec:sim}
Based on the real-world Twitter network \citep{guess2016treatments},
we conduct a simulation study where we generate a variety of
unobserved offline face-to-face networks. We evaluate how well the
sensitivity analysis methods estimate the Twitter-specific spillover
effect. 

We vary the simulation setup along two dimensions; (1) the overlap between
the observed Twitter network and a simulated unobserved offline
network and (2) the outcome data generating process --- a linear
additive model or an interactive model. With this design, we 
illustrate three key results we derived analytically. First, the
bias increases when the overlap
between the online network and the unobserved offline  
network increases (Theorem~\ref{biasASE}). Second, the parametric sensitivity analysis can
recover unbiased estimates under a linear additive model but suffers from bias
under an interactive model (Theorem~\ref{Parasen1}).  Finally, the
nonparametric sensitivity analysis provides bounds for the ANSE under
both models, but it has larger confidence intervals than the parametric
version under the linear additive model (Theorem~\ref{sen2}).

\paragraph{Background.}
To make our simulation as realistic as possible, we rely on the
real-world Twitter network studied in
\cite{guess2016treatments}. The original authors conducted 
a political mobilization experiment over the Twitter network to
estimate the effectiveness of online mobilization appeals. In
particular, they sampled followers of the Twitter account
of a nonprofit advocacy organization, the League of Conservation
Voters (LCV), and measured a network adjacency matrix among them. 
Each node of the network is a Twitter user who follows the
LCV’s account and a directed edge exists from user $i$ to user $j$ when
user $i$ follows user $j$. Instead of artificial simulated networks,
we use this real-world Twitter network as the basis of our simulations
after preprocessing as described below.

\paragraph{Simulation Design.}
To make the comparison of sensitivity analysis methods clear, we 
preprocess the Twitter network of
\cite{guess2016treatments}. First, to avoid the influence of outliers, we remove units that
are above the 95 percentile of the distribution on the size of Twitter
neighbors (also known as the out-degree distribution). To have
well-defined treated proportions $G_i$ (explain more below), we also remove those who follow less than
five other units. The resulting network contains $2430$ units with the
mean number of neighbors equal to $18.5$. We simulate an unobserved
offline network $\cU$ with four different values 
of the overlap $\pi_{GU}\in\{0.0, 0.2, 0.4, 0.6\}$ where we expect
no bias in the case of $\pi_{GU} = 0$. We use the Bernoulli design to randomly assign
treatments with probability $0.5.$ 

We consider two outcome data generating
processes. For the linear additive model, we generate potential
outcomes for individual $i$ by $ Y_i (T_i, G_i, U_i) \ = 5 + 2 T_i + G_i + 1.5 U_i + \epsilon_i,$
where  $\epsilon_i$ is randomly drawn from a normal
distribution with $(\mu,  \sigma)  = (0.0, 0.5)$. $G_i$ and $U_i$ denote treated proportions of neighbors
in the observed Twitter network and unobserved treated proportions of
neighbors in the offline network, respectively. For the interactive
model, we use $ Y_i (T_i, G_i, U_i) \ = 5 + 2 T_i + G_i +
2 T_i \times U_i + 2 G_i \times U_i + 0.5 U_i + \epsilon_i$. As the
main causal estimand, we consider the ANSE, $\ANSE(0.6, 0.4; 0)$, where we compare $g^H=0.6$ relative to
$g^L=0.4$ when $d=0$ (to have $G_i = 0.6$ and $G_i= 0.4$ both 
well-defined, we focus on a subset of units who follows at least five
units). In the linear additive model, the true ANSE is equal to $0.2$, and in
the interactive model, the true ANSE is approximately equal to $0.4$
(the exact values change according to the overlap $\pi_{GU}$).

The offline network $\cU$ is unmeasured and thus, we can rely only on
a linear regression $Y_i  \sim \alpha + \beta T_i + \gamma_0 G_i + \gamma_1 (T_i \times G_i)$ with
weights $w_i^{\texttt{B}} = 1/\Pr(T_i, G_i).$ An estimator,
$0.2 \times \widehat{\gamma}_0,$ is biased for $\ANSE(0.6, 0.4; 0)$ due to unobserved network $\cU.$ The
parametric sensitivity analysis provides a bias-corrected estimate,
$0.2 \times \widehat{\gamma}_0 -  0.2 \times \lambda \times \pi_{GU},$
with two sensitivity parameters; the overlap $\pi_{GU}$ and the
spillover effect in the unobserved network $\lambda$.\footnote{\spacingset{1}{\footnotesize $\pi_{GU} \in \{0.0, 0.2, 0.4,
  0.6\}$. In the linear additive model case, $\lambda =
  1.5$. In the interactive model case, while $\lambda$ is not
  well-defined, we use the main effect of
  $U_i$ ($0.5$) as an example and show that the parametric sensitivity analysis
  cannot remove bias due to the violation of the linear additive
  assumption.}} The nonparametric sensitivity analysis computes bounds with sensitivity
parameters $(\MR_{UY}, \RR_{GU})$.\footnote{\spacingset{1}{\footnotesize Imputing the potential
    outcome model into the definition, $\MR_{UY} = 1.16$ (linear
    additive model)
    and $\MR_{UY} = 1.14$ (interactive model). Using the definition of
    the overlap, $\RR_{GU} = (0.6 \times \pi_{GU} +  0.5\times (1-\pi_{GU}))/(0.4 \times \pi_{GU} +
    0.5\times (1-\pi_{GU}))$ (both models).}} We evaluate an estimator
that ignores unobserved offline networks, the parametric sensitivity
analysis, and the nonparametric sensitivity analysis with the $2000$
Monte Carlo simulations.  

\begin{figure}[!t]
  \begin{center}
    \includegraphics[width = \textwidth]{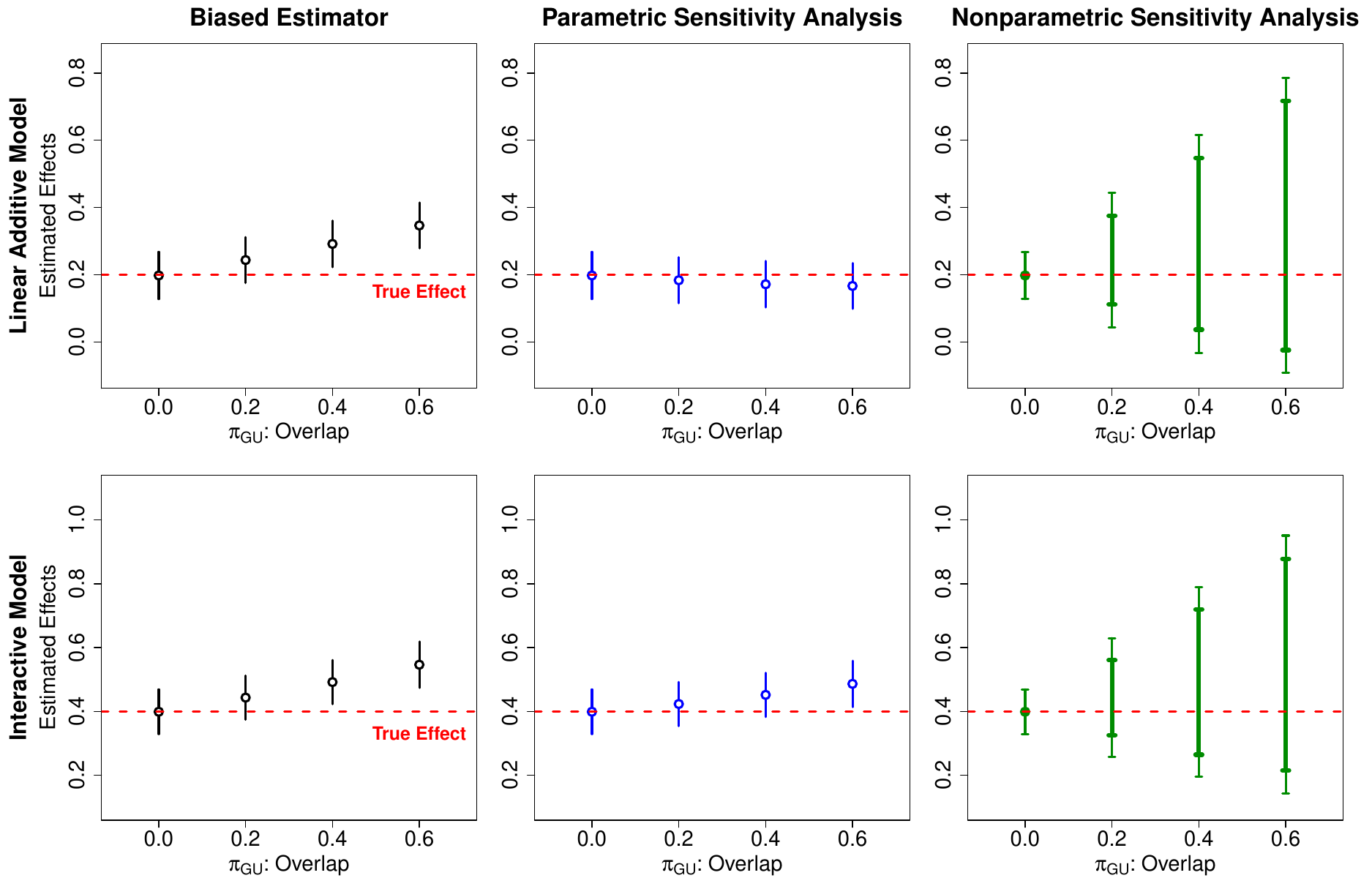}
  \end{center}      
  \spacingset{1}{
    \caption{Simulation Results on Bias and Sensitivity Analysis for the Twitter Experiment. {\it Note}: The first
      (second) row reports results from the linear additive (interactive)
      model. The first column demonstrates the bias in estimators that ignore the
      unobserved offline network. The second column shows that the
      parametric sensitivity analysis can recover unbiased estimates
      of the ANSE under the linear additive model but not under the interactive
      model. The third column demonstrates that bounds from the
      nonparametric sensitivity analysis cover the true ANSE, but are
      much wider than the parametric version under the linear additive
      model.}\label{fig:bias}
  }
\end{figure}

\paragraph{Results.}
Figure~\ref{fig:bias} presents simulation results. The first column
reports estimates, $0.2 \times \widehat{\gamma}_0,$ that ignore the
unobserved network $\cU$, with the 95\% confidence intervals. 
Both settings of the linear additive and interactive models
illustrate that the estimator is biased unless the overlap
$\pi_{GU}$ is zero, and the bias increases as the overlap increases. The second column shows the results for the parametric
sensitivity analysis with the 95\% confidence intervals. In the linear
additive model setup (the first row), the
parametric sensitivity analysis recovers approximately unbiased
estimates. In contrast, in the interactive model setup (the second
row), it still suffers from bias as the linear additive assumption
(Assumption~\ref{linear}) is violated. Finally, the third column
reports bounds from the nonparametric sensitivity analysis  (thick
green bars) and the 95\% confidence intervals of the bounds (thin
green bars). When the overlap $\pi_{GU}$ is zero, the bound converges
to an unbiased point estimate of the ANSE. Importantly, the bounds cover the true ANSE in both
linear additive and interactive model settings as Theorem~\ref{sen2} implies. It
also reveals an important limitation; while the nonparametric bounds
cover the true ANSE, they are much wider than the 95\% confidence intervals
of the parametric sensitivity analysis under the linear additive model. This is
the case especially  when the overlap between the main network of interest and the unobserved network is large. 

\section{Empirical Application: Network Field Experiment}
\label{sec:app}
\cite{cai2015social} are interested in understanding how farmers in rural China use 
social networks to make important financial decisions, i.e., insurance
take-up. They designed an experiment with
China's largest insurance provider, the People's Insurance Company of
China (PICC), to randomly assign households to different information
sessions about the insurance. While they estimate the direct treatment
effect as a first step, the primary focus of the original 
analysis is to estimate the spillover effect of such information on
the insurance take-up.

As in many other network field experiments, they use a social network
survey to ask experimental subjects to name their network
connections. In particular, they asked household heads to name five
close friends with whom ``they most frequently discuss rice production
or financial issues'' \citep[p. 88]{cai2015social}. Although this is a
popular strategy, there are several reasons to afraid of unobserved
networks. First, as carefully noted in the original paper, almost all
households listed the maximum number of friends (the average number of
listed friends is $4.9$ where respondents can only list up to $5$
friends), which suggests that respondents could have listed more
friends if other forms of network surveys were used \citep{larson2019measuring}. Second, networks
among experimental subjects are dense; ``the average path-length is
about $2.67$, which means that a household can be connected to any other
household in the village by passing on average of two to three
households'' \citep[p. 89]{cai2015social}. This suggests that there are
many potential network connections through which experimental subjects
can communicate with each other. For example, in such rural Chinese
villages, a kinship network is of great importance \citep[e.g.,][]{xiong2017characteristics}.  

We extend the original analysis by estimating the
spillover effect specific to the observed financial network while
accounting for unobserved networks, such as the kinship network. 

\paragraph{Setup.} 
\cite{cai2015social} designed the experiment with two rounds to
estimate the spillover effect. 2175 households were invited to the
first round and they were randomly assigned into one of two information
sessions, \texttt{simple} or \texttt{intensive}. The \texttt{simple}
session took about 20 minutes and the PICC agents explained the
insurance contract. The \texttt{intensive} sessions took about 45
minutes and provided all the information given in the \texttt{simple}
session, plus an additional detailed explanation of expected benefits
and costs of purchasing the insurance. Three days after the first
round, a different set of households were invited to the second round
and randomly assigned into the \texttt{simple} or \texttt{intensive}
sessions. We follow the original analysis and focus on 1317
households who were invited to this second round and only received
information from the \texttt{simple} or \texttt{intensive}
session (657 and  660 households,
respectively).\footnote{\spacingset{1}{\footnotesize In the original
    paper, these groups are labeled 
    as Simple2-NoInfo and Intensive2-NoInfo. There were two other randomly assigned groups
    who received additional information about the take-up
    rates in the first round. The original analysis 
    (Table 2 in the original paper) focuses on the first two groups,
    which we follow.}} The main outcome of interest is a binary
variable whether each household in the second round takes up the
insurance. Table~\ref{tab:ex} summarizes the relevant aspects of the experimental
design. See \cite{cai2015social} for details and other features of the
design. 

For participants in the second round, the original authors define the
main exposure variable of interest $G_i$ to be the proportion of peers in
their financial network who attended the first round
\texttt{intensive} session. The direct treatment $T_i$ is defined as
an indicator variable taking $1$ if household $i$ receives the
\texttt{intensive} session and $0$ otherwise. Following their
analysis, we focus on the ANSE specific to the observed financial
network and compare $g^H = 0.2$ and $g^L =  0$ under no direct
treatment receipt $d= 0$, formally $\ANSE(g^H = 0.2, g^L =  0;  d =
0)$, as they show the biggest difference is between no treated peer
and one treated peer (Table 3 of the original paper). In addition, to satisfy
the standard overlap assumption \citep{imbens2015causal}, we analyze units who listed the
maximum number of five households\footnote{\spacingset{1}{\footnotesize If the size of neighbors is
smaller than five, $g^H = 0.2$ is impossible for such units, which
violates the overlap assumption.}} in the financial network question
(1207 households, more than 90\% of all the samples). 

\begin{table}[!t]
\centering \small
\begin{tabular}{|cc|cc|}
  \hline
  \multicolumn{2}{|c|}{First round} & \multicolumn{2}{c|}{Second round}\\ 
  \hline
  \texttt{simple}  & \texttt{intensive} & \texttt{simple}  & \texttt{intensive} \\ 
  \hline 
  1079 & 1096 & 657 & 660\\ 
  \hdashline
  \multicolumn{2}{|c|}{total: 2175} & \multicolumn{2}{c|}{total: 1317}\\ 
  \hline
\end{tabular}
\caption{Experimental Design.}\label{tab:ex}
\end{table}

To estimate spillover effects, the original authors estimate the
following linear regression (model (4) in Table 2 of
\cite{cai2015social}) with standard errors clustered at the village level. 
\begin{equation}
  Y_i = \alpha + \beta T_i + \gamma_0 G_i + \gamma_1 (T_i \times G_i) + \bX^\top_i \theta + \epsilon_i
\end{equation}
where pre-treatment covariates $\bX_i$ include the size of neighbors, village
fixed effects, and other household characteristics (gender, age,
education of household head, rice production area, risk aversion, and
perceived probability of future disaster). 
Thus, if the observed financial network is the only causally relevant network (i.e., the no
omitted network assumption holds), an estimator 
$0.2 \times \widehat{\gamma}_0$ is unbiased for the ANSE,
$\ANSE(g^H=0.2, g^L=0;d=0)$ where $\widehat{\gamma}_0 = 0.444$ 
(s.e. =  $0.109$) in the original paper.
However, as studied in Section~\ref{sec:bias},
when the no omitted network assumption is violated --- for example,
people might share information through their kinship network, the
estimator is biased for the ANSE specific to the financial network. 

\paragraph{Sensitivity Analysis.} 
We use the proposed sensitivity analysis methods to investigate the
robustness of the original findings to unobserved networks. First, for
the parametric sensitivity analysis, we consider three values for each
of the two sensitivity parameters, the spillover effect in the
unobserved network $\lambda \in (0.1, 0.2, 0.4)$ and
the overlap $\pi_{GU} \in (0.2, 0.4, 0.6)$, producing a total of
nine scenarios. The three values of $\lambda$ are chosen to represent
different scenarios in which we expect spillover effects in the kinship network is much smaller than,
smaller than, or similar to the total spillover effect in the
financial network ($\widehat{\gamma}_0 = 0.444$). Similarly, the three
values of $\pi_{GU}$ cover from small to large overlaps. For the
nonparametric sensitivity analysis, we also investigate three values for each
of the two sensitivity parameters $\MR_{UY} \in (1.3, 1.5, 1.8)$ and
the overlap $\RR_{GU} \in (1.3, 1.5, 1.8)$. Although these are
relatively small risk and mean ratios (e.g., see Table 1 in
\cite{ding2016sensitivity}), we see below that substantive conclusions
change according to different sensitivity parameters. Finally, while the original paper uses a linear regression without weights, 
we use a weighted linear regression with weights $w_i^{\texttt{B}} = 1/\Pr(T_i, G_i)$ as proposed in Section~\ref{sec:sen} for both
analyses.

\begin{figure}[!t]
  \begin{center}
    \includegraphics[width = \textwidth]{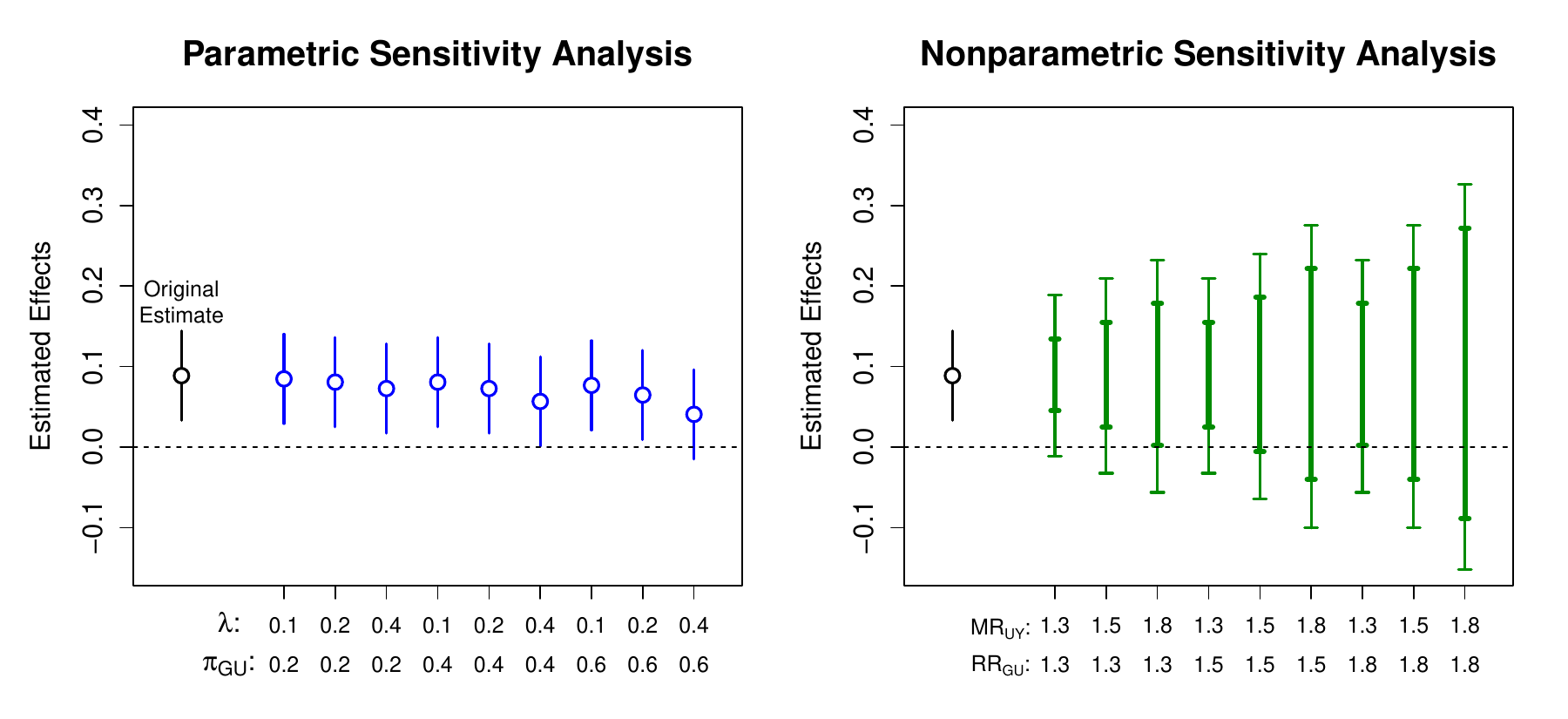}
  \end{center}  \vspace{-0.25in}    
  \spacingset{1}{
    \caption{Parametric and Nonparametric Sensitivity Analysis for
      the Field Network Experiment in China. {\it Note}: The left
      (right) panel reports the parametric (nonparametric) sensitivity
      analysis with different sensitivity parameters. In both panels, the
      first black point shows an estimate of the ANSE that ignores
      unobserved networks. In
      the left panel, the nine other blue points represent estimates from the parametric sensitivity analysis
      with their 95\% confidence intervals. In the right panel, the
      thick green bars represent
      nonparametric bounds on the ANSE and the thin green bars are their
      95\% confidence intervals. }\label{fig:china}
    }
\end{figure}

Figure~\ref{fig:china} presents results from the parametric (left) and
nonparametric sensitivity analysis (right). In the left panel, the
first black point shows an estimate of the ANSE that ignores the unobserved kinship network with its 95\% confidence interval ($8.87$
percentage point; 95\% CI = $[3.31, 14.42]$). The nine other blue
points represent estimates from the parametric sensitivity analysis
with their 95\% confidence intervals. Importantly, point estimates are
relatively stable over a range of sensitivity parameters and the 95\%
confidence interval covers zero only when the effect in the
unobserved kinship network is large ($\lambda = 0.4$) and the overlap
between the observed financial network and the unobserved kinship
network is relatively large ($\pi_{GU} = 0.6$). Thus, the
parametric sensitivity analysis suggests that, while point estimates of the
ANSE specific to the observed financial network are smaller than the
original estimates, estimates are still positive and statistically
significant at the conventional level of $0.05$. Although this
parametric sensitivity analysis is simple and intuitive, it requires the
parametric linear additive assumption as discussed in
Section~\ref{subsec:parasen}. We now turn to the nonparametric
sensitivity analysis which requires weaker assumptions at the expense of
efficiency. 

The right panel of Figure~\ref{fig:china} shows results of the
nonparametric sensitivity analysis where the first black point
reproduces the same ANSE estimate that ignores the unobserved kinship
network as a reference point. The thick green bars represent
nonparametric bounds on the ANSE and the thin green bars are their
95\% confidence intervals. Although lower bounds are positive for
small bias scenarios, such as  $(MR_{UY}, RR_{GU}) = (1.3, 1.3), (1.5,
1.3), (1.3, 1.5),$ all the 95\% confidence intervals are inconclusive
about the sign of the ANSE. Importantly, this result reveals that the
positive ANSE estimates are sensitive to different assumptions about unobserved networks in the
nonparametric sensitivity analysis, in contrast to the parametric sensitivity analysis above. 

This difference in results from the parametric and nonparametric
sensitivity analysis can arise for two reasons. First, the linear
additive assumption (Assumption~\ref{linear}) required for the parametric
sensitivity analysis could be violated and therefore, the nonparametric
bounds are more credible. Second, the nonparametric bounds are in
general less efficient than parametric methods and thus, large
nonparametric bounds simply indicate insufficient statistical power
for detecting the ANSE non-parametrically. In this China experiment, as
the original authors detect the non-linearity in the total spillover effect
in the financial network (Table 3 of \cite{cai2015social}), it is more
likely that the linear additive assumption is violated and thus, the
nonparametric sensitivity analysis is more credible. In practice, it is important
to conduct both parametric and nonparametric sensitivity analysis
methods as they are complementary and evaluate the robustness of the ANSE estimates under different assumptions.

\section{Concluding Remarks}
\label{sec:con}
Although early work in the causal inference literature has assumed no
interference, a growing number of both applied and methodological
papers now explicitly incorporate spillover effects to understand
causal interdependence across people. In this paper, we propose a
framework for analyzing spillover effects in common social science
settings of multiple networks, and introduce a new causal estimand, the average network-specific spillover effect
(ANSE), to separately quantify the amount of spillover effects in each
network. We show that the unbiased estimation of the ANSE requires an often-violated
assumption of no omitted network. Unlike conventional omitted variable
bias, the bias due to unobserved networks remains even when treatment assignment is randomized and when the network of interest
and unobserved networks are independently generated. To account for
this bias, we provide parametric and nonparametric sensitivity analysis
methods, with which researchers can assess the robustness of causal
conclusions to unobserved networks. The proposed methods are
illustrated by two common types of network experiments; an online social network experiment and a network field experiment.

There are a number of future extensions that can further improve the
methodologies proposed in this paper. First, although we made the
assumption of stratified interference throughout this paper, we can
potentially derive the exact bias formula and sensitivity analysis
methods without it. This direction will be particularly important
since not only the no omitted network assumption but also the
assumption of stratified interference might be strong in many applied
settings. Second, while we focused on the estimation of the ANSE in
this paper, it would also be of significant interest to study how to
incorporate observed and unobserved networks into the rich literature
of the Fisher randomization test for interference
\citep[e.g.,][]{rosenbaum2007, aronow2012random, bowers2013interference, athey2016exact,
   basse2019}. Third, it is useful to study 
how we can effectively incorporate a supplementary survey to estimate the overlap between the main network of
interest and an unobserved network. For example, in an online network experiment, even though it might
be difficult to measure an offline face-to-face network for every
subject in the experiment, researchers can use a network survey method \citep[e.g.,][]{bisbee2017testing}
to randomly sample subjects and use the estimated overlap for the
sensitivity analysis. Fourth, this paper primarily focused on the bias
due to unobserved networks, but another important extension would be to study
implications of omitted relevant networks to variance estimation. A
promising approach would be to consider a range of sensitivity parameters
and compute the worst-case confidence interval \citep[e.g.,][]{berger1994p, aronow2016confidence}. Finally, as discussed
in the paper, the estimation accuracy is often of concern in settings with
multiple networks. Thus, it would be useful to extend the literature
on experimental design for spillover effects to settings of  multiple
networks and study how to design the optimal experiment
\citep[e.g.,][]{hudgens2008toward, sinclair2012detecting, basse2018model,
  bowers2018models, jagadeesan2019designs}.

\clearpage
\vspace{0.1in}
\spacingset{1.0}
\pdfbookmark[1]{References}{References}
\bibliography{egami}

\clearpage
\appendix
\spacingset{1.25}
\setcounter{table}{0}
\setcounter{equation}{0}
\setcounter{figure}{0}
\renewcommand {\theequation} {A\arabic{equation}}
\renewcommand {\thetable} {A\arabic{table}}
\renewcommand {\thefigure} {A\arabic{figure}}

\begin{center}
  \spacingset{1.5}
  {\LARGE Supplementary Appendix for: \\ Naoki
    Egami. ``Spillover Effects in the
      Presence \\ of Unobserved Networks.''
    {\it Political Analysis}}
  \setcounter{page}{1}
\end{center}


  \spacingset{1.45}
  \section{Details of Setup}
  \label{sec:regular}
  We describe regularity conditions for the support of treatment
  exposure probabilities to ensure well-defined causal estimands.

  The required regularity conditions are as follows: (1) the support of
  $\Pr(G_i=g, U_i=u | T_i=1)$ is equal to the support of $\Pr(G_i=g, U_i=u | T_i=0)$ for all $i$,
  and (2) the support of $\Pr(U_i=u | T_i=d, G_i=\ag)$ is equal to
  the support of $\Pr(U_i=u | T_i=d, G_i=\abg)$ for all $i$. We discuss them in order. 

  When we define the unit level direct effect, we avoid ill-defined causal
  effects by focusing on settings where the support of $\Pr(G_i=g, U_i=u \mid T_i=1)$ is
  equal to the support of $\Pr(G_i=g, U_i=u \mid T_i=0)$ for all $i$. 
  This can be violated when the total number of treated units is
  small so that for some $(g, u)$, $\Pr(G=g, U=u | T_i=1) =0$ and
  $\Pr(G=g, U=u | T_i=0) >0.$ One extreme example is that when we use
  complete randomization with the total number of treated units equal to
  $1$. In this case, whenever $T_i=1$, $\Pr(G=g, U=u | T_i=1) =0$ for
  all $(g,u)$, but when $T_i=0$, $\Pr(G=g, U=u | T_i=0) >0$ for
  some $(g,u).$ Another extreme example is that the total number of
  treated units is too large. For example, when we use complete
  randomization with the total number of treated units equal to $N-1$. 
  In this case, whenever $T_i=0$, $\Pr(G=g, U=u | T_i=0) =0$ for
  all $(g,u)$ except for $(g, u) = (1,1)$, but when $T_i=1$, $\Pr(G=g, U=u | T_i=1) >0$ for
  some $(g,u)$ other than $(g, u) = (1,1)$. It is clear that 
  when researchers use a Bernoulli design, the support of $\Pr(G_i=g, U_i=u \mid T_i=1)$ is
  equal to the support of $\Pr(G_i=g, U_i=u \mid T_i=0)$ for all $i$. 

  When we define the unit level network-specific spillover effect, we avoid ill-defined causal
  effects by focusing on settings where $\Pr(U_i=u \mid T_i=d, G_i=\ag)$
  and $\Pr(U_i=u \mid T_i=d, G_i=\abg)$ have the
  same support for all $i$. This requires that $g^H$ and $g^L$ are
  small enough so that the distribution over the fraction of treated neighbors in network
  $\cU$ is not restricted, especially $\Pr(U_i=0 \mid T_i=d, G_i=\ag) > 0$
  and $\Pr(U_i=0 \mid T_i=d, G_i=\abg) > 0$ for all $i$. 
  Formally, $\ag, \abg \leq g_s$ where $g_s \equiv \underset{i}{\min} \{1- |\cN^{(\cG,\cU)}_i|/|\cN^{\cG}_i|\}$. 
  The desired support condition can be violated when the total number of
  treated units is too small so that for some $u$, $\Pr(U=u | T_i=1, G_i=\ag) =0$ and
  $\Pr(U=u | T_i=1, G_i=\abg) >0$. One extreme example is that when we use
  complete randomization with the total number of treated units equal to
  $1 + \ag \times |\cN^{\cG}_i|.$ In this case, whenever $G_i=\ag$,
  $\Pr(U=u | T_i=1, G_i=\ag) =0$ for all $u > |\cN^{(\cG,\cU)}_i|/|\cN^{\cU}_i|$, 
  but when $G_i=\abg < \ag$, $\Pr(U=u | T_i=0, G_i=\abg) >0$ for
  some $u > |\cN^{(\cG,\cU)}_i|/|\cN^{\cU}_i|$. Finally, it is clear
  that $\Pr(U_i=u \mid T_i=d, G_i=\ag)$ and $\Pr(U_i=u \mid T_i=d,
  G_i=\abg)$ have the same support for all $i$ if researchers use a
  Bernoulli design and $\ag, \abg \leq g_s$. 

\section{Connection between Total Spillover Effects and Network-Specific Spillover Effects}
\label{sec:decom}
Here, we connect the ANSE to the popular estimand in the
literature. In particular, we show that the ANSE can be seen as the
decomposition of the average total spillover effect
\citep{hudgens2008toward}. 

First, by extending \cite{hudgens2008toward} to settings with multiple
networks, the individual average potential outcome are defined as follows.
\begin{equation}
\overline{Y}_i (d,g) \equiv \sum_{u \in \Delta_i^u} Y_i(d, g, u) \Pr(U_i=u \mid T_i=d, G_i=g),
\end{equation}
where the potential outcome of individual $i$ is averaged over the
conditional distribution of the treatment assignment $\Pr(U_i=u \mid
T_i=d, G_i=g).$ Here, the individual average potential outcome
represents the expected outcome of unit $i$ when she receives the
direct treatment $d$ and the treated proportion $g$ in network $\cG$. Taking the difference in the two individual
average potential outcomes, the {\it average total spillover effect}
(ATSE) in network $\cG$ is defined as follows
\citep{halloran2016review}.\footnote{This quantity is called the
  average indirect causal effect in \cite{hudgens2008toward}. We define it as
  the average total spillover effect to clarify how it combines
  multiple network-specific spillover effects.}
\begin{equation}
  \ATSE(\ag, \abg;d) \ \equiv \  \frac{1}{N} \sum_{i=1}^N \{\overline{Y}_i (d, \ag) - \overline{Y}_i (d, \abg)\}.
\end{equation}
This causal quantity is the {\it total} spillover effect of changing
the treated proportion in network $\cG$ from $g^L$ to $g^H$ as the
following decomposition of the ATSE demonstrates.
\begin{eqnarray}
  \ATSE(\ag, \abg;d) & = &  \ANSE(\ag, \abg;d) +  \label{eq:decom}\\[5pt]
  && \hspace{-1.8in}\frac{1}{N} \sum_{i=1}^N \l\{\sum_{u \in \Delta^u_i} \{Y_i(d, \ag, u)
     -Y_i(d, \ag,u^\prime)\} \{\Pr (U_i=u \mid T_i=d, G_i=\ag) -
     \Pr(U_i=u  \mid T_i=d, G_i=\abg)\} \r\},  \nonumber
\end{eqnarray}
for any $u^\prime \in \Delta_i^u$. The first term is the ANSE in
network $\cG$ (Definition~\ref{def:ANSE}), which
quantifies the spillover effect specific to network $\cG$. The second term
represents the spillover effect in $\cU$, $Y_i(d, \ag, u) -Y_i(d, \ag,u^\prime)$, weighted by the
change in the conditional distribution of $U_i$ due to the change in $G_i$,  $\Pr (U_i=u \mid T_i=d, G_i=\ag) - \Pr(U_i=u  \mid T_i=d,
G_i=\abg).$ This is because $U_i$, the treated proportion of neighbors in the
other network $\cU$, is not fixed at constant and thus, they change as
$G_i$, the treated proportion of neighbors in $\cG$, changes. Thus, the ATSE captures the sum of the 
spillover effect specific to network $\cG$ and the
spillover effect specific to $\cU$ induced by the change in $U_i$
associated with the change in $G_i$. For example, the ATSE of changing
from $g^L$ to $g^H$ on the Facebook network captures two
spillover effects together; (1) the spillover effect specific to the
Facebook and (2) the spillover effect in the face-to-face network. This is because the treated proportion in the offline network
$U_i$ is associated with the change in the treated proportion in the
Facebook network $G_i$. We discuss this issue in further details when
we derive the exact bias formula in Section~\ref{sec:bias}. When
network $\cU$, such as the offline network, is causally irrelevant,
the ATSE is equal to the ANSE in the Facebook network, but in general,
the two estimands do not coincide.

While both the ATSE and the ANSE quantify spillover effects, their
substantive meanings differ. The ATSE is useful when researchers wish to know the total amount of
spillover effects that result from interventions on an observed network. 
For instance, politicians decided to run online campaigns on Twitter
and want to estimate the total amount of spillover effects they can
induce by their Twitter messages. These politicians might not be
interested in distinguishing whether the spillover effects arise through Twitter
or through unobserved face-to-face interactions. Thus, the ATSE is of
relevance when the target network is predetermined and the mechanism can be ignored. 

In contrast, the ANSE is essential for disentangling
different channels through which spillover effects arise. 
It is the main quantity of interest when researchers wish
to examine the causal role of individual networks or to discover the
most causally relevant network to target. For example, it is of
scientific interest to distinguish how much spillover effects arise
through the Twitter network or through offline communications. By estimating the ANSE, researchers can
learn about the importance of online human interactions.

  \section{Proofs}
  This section provides proofs for all theorems in the paper. 
  \subsection{Proof of Theorem~\ref{estimation}}
  \label{app:identification}

  \subsubsection{ADE}

  First, we rewrite the estimator with the standard IPW representation.
  \begin{eqnarray*}
    && \widehat{\ADE} \\
    & = & \frac{1}{N} \sum_{i=1}^N \mo\{T_i=1\}  \widetilde{w}_i Y_i - \frac{1}{N} \sum_{i=1}^N \mo\{T_i=0\} \widetilde{w}_i Y_i \\
    & = &  \frac{1}{N} \sum_{i=1}^N \sum_{(g,u) \in \Delta^{gu}_i} \Pr (G_i=g, U_i=u) 
          \biggl\{ \frac{\mo\{T_i=1, G_i=g,
          U_i=u\} Y_i}{\Pr(T_i=1, G_i=g, U_i=u)} - \frac{\mo\{T_i=0, G_i=g,
          U_i=u\} Y_i}{\Pr(T_i=0, G_i=g, U_i=u)}\biggr\}.
  \end{eqnarray*}
  Then, the theorem follows from the standard proof for the IPW estimator. 
  \begin{eqnarray*}
    && \E[\hat{\ADE}]  \\
    & = & \frac{1}{N} \sum_{i=1}^N
          \sum_{(g,u) \in \Delta^{gu}_i} \Pr
          (G_i=g, U_i=u) \times \\
    && \hspace{1in} \biggl\{ \frac{\E[\mo\{T_i=1, G_i=g,
       U_i=u\} Y_i]}{\Pr(T_i=1, G_i=g, U_i=u)} - \frac{\E[\mo\{T_i=0, G_i=g,
       U_i=u\} Y_i]}{\Pr(T_i=0, G_i=g, U_i=u)}
       \biggr\} \\
    & = & \frac{1}{N} \sum_{i=1}^N
          \sum_{(g,u) \in \Delta^{gu}_i} \Pr
          (G_i=g, U_i=u) \times \\
    && \hspace{1in} \biggl\{ \frac{\Pr(T_i=1, G_i=g,
       U_i=u) Y_i(1, g, u)}{\Pr(T_i=1, G_i=g, U_i=u)} - \frac{\Pr(T_i=0, G_i=g,
       U_i=u) Y_i(0,g,u)}{\Pr(T_i=0, G_i=g, U_i=u)}
       \biggr\} \\
    & = & \frac{1}{N} \sum_{i=1}^N \sum_{(g,u) \in \Delta^{gu}_i} \Pr (G_i=g, U_i=u) \{ Y_i(1, g, u) - Y_i(0,g,u) \} \\
    & = & \ADE
  \end{eqnarray*}
  where the second equality follows from the consistency of potential outcomes. \qed

  \subsubsection{ANSE}

  First, we rewrite the estimator with the standard IPW representation.
  \begin{eqnarray*}
    && \widehat{\ANSE} \\
    &  = &  \frac{1}{N} \sum_{i=1}^N \mo\{T_i = d, G_i = \ag\}
           w_i Y_i - \frac{1}{N} \sum_{i=1}^N \mo\{T_i = d, G_i = \abg\} w_i Y_i \\
    & = &  \frac{1}{N} \sum_{i=1}^N \sum_{u \in \Delta_i^u} \Pr
          (U_i=u \mid T_i=d, G_i = \abg)
          \times \\ 
    && \hspace{1.5in} \biggl\{ \frac{\mo\{T_i=d, G_i=\ag,
       U_i=u\} Y_i}{\Pr(T_i=d, G_i=\ag, U_i=u)} - \frac{\mo\{T_i=d,
       G_i=\abg, U_i=u\} Y_i}{\Pr(T_i=d, G_i= \abg, U_i=u)}
       \biggr\},
  \end{eqnarray*}
  Then, the theorem follows from the standard proof for the IPW estimator. 
  \begin{eqnarray*}
    && \E[\hat{\ANSE}(g,g^\prime; d)]  \\
    & = & \frac{1}{N} \sum_{i=1}^N \sum_{u \in \Delta_i^u} \Pr
          (U_i=u \mid T_i=d, G_i = \abg) \times \\ 
    && \hspace{1in} \biggl\{ \frac{\E[\mo\{T_i=d, G_i=\ag,
       U_i=u\} Y_i]}{\Pr(T_i=d, G_i=\ag, U_i=u)} - \frac{\E[\mo\{T_i=d, G_i=\abg,
       U_i=u\} Y_i]}{\Pr(T_i=d, G_i=\abg, U_i=u)}
       \biggr\} \\
    & = & \frac{1}{N} \sum_{i=1}^N \sum_{u \in \Delta_i^u} \Pr
          (U_i=u \mid T_i=d, G_i = \abg) \times \\ 
    && \hspace{0.5in} \biggl\{ \frac{\Pr (T_i=d, G_i=\ag,
       U_i=u) Y_i(d,\ag,u)}{\Pr(T_i=d, G_i=\ag, U_i=u)} - \frac{\Pr (T_i=d, G_i=\abg,
       U_i=u) Y_i(d, \abg,u)}{\Pr (T_i=d, G_i=\abg, U_i=u)}
       \biggr\} \\
    &  = & \frac{1}{N} \sum_{i=1}^N \sum_{u \in \Delta_i^u} \Pr
           (U_i=u \mid T_i=d, G_i =  \abg) \{Y_i(d,\ag,u) - Y_i(d,\abg,u)\} \\
    & = & \ANSE (\ag,\abg; d),
  \end{eqnarray*}
  which completes the proof.  \qed

\subsection{Proof of Theorem~\ref{biasASE}}
  \label{app:biasASE}
  The expectation of an estimator $\hat{\ANSE}_B(\ag, \abg; d)$ is 
  \begin{eqnarray*}
    && \E[\hat{\ANSE}_B(\ag, \abg; d)]   =  \frac{1}{N} \sum_{i=1}^N 
       \biggl\{ \E[Y_i \mid T_i = d, G_i =  g^H] - \E[Y_i \mid T_i = d, G_i =  g^L] \biggr\} \\
    & = &  \frac{1}{N} \sum_{i=1}^N \sum_{u \in \Delta^u_i} \biggl\{
          Y_i (d,\ag,u) \Pr(U_i=u \mid T_i=d, G_i=\ag) -
          Y_i(d, \abg,u) \Pr(U_i=u \mid T_i=d, G_i=\abg) \biggr\}.
  \end{eqnarray*}
  Therefore, we get  
  \begin{eqnarray*}
    && \E[\hat{\ANSE}_B(\ag, \abg; d)]   - \ANSE(\ag, \abg; d) \\
    & = &  \frac{1}{N} \sum_{i=1}^N \sum_{u \in \Delta^u_i} \biggl\{
          Y_i (d,\ag,u) \{\Pr(U_i=u \mid T_i=d, G_i=\ag) - \Pr(U_i=u \mid
          T_i=d, G_i = \abg)\} \\
    &&  \qquad - Y_i(d, \abg,u) \{\Pr(U_i=u \mid T_i=d, G_i=\abg)
       - \Pr(U_i=u \mid T_i=d, G_i = \abg)\} \biggr\} \\
    & = &  \frac{1}{N} \sum_{i=1}^N \sum_{u \in \Delta^u_i} \biggl\{
          \{Y_i (d, \ag,u) - Y_i(d, \ag, u^\prime)\} \\
&& \hspace{1.2in} \times \{\Pr(U_i=u \mid T_i=d, G_i=\ag) - \Pr(U_i=u \mid
          T_i=d, G_i = \abg)\} \biggr\}.
  \end{eqnarray*}
  for any $u^\prime \in \Delta^u$. 
  \qed

  \subsubsection{Lemma: Bias in ADE}
  \label{app:biasADE}
  First, we can rewrite the estimator as the standard IPW estimator. 
  \begin{eqnarray*}
    \widehat{\ADE}_B   &  = & \frac{1}{N} \sum_{i=1}^N \mo\{T_i=1\}
                              \widetilde{w}^{\texttt{B}}_i Y_i - \frac{1}{N} \sum_{i=1}^N \mo\{T_i=0\}
                              \widetilde{w}^{\texttt{B}}_i Y_i  \\
                       &  = &  \frac{1}{N} \sum_{i=1}^N  \sum_{g \in \Delta^{g}_i} 
                              \Pr (G_i=g) \biggl\{ \frac{\mo\{T_i=1, G_i=g\} Y_i}{\Pr(T_i=1,
                              G_i=g)} -  \frac{\mo\{T_i=0, G_i=g\} Y_i}{\Pr(T_i=0,
                              G_i=g)} \biggr\}.
  \end{eqnarray*}
  We have the following equality for any $g$,
  \begin{eqnarray*}
    && \E[\mo\{T_i=d, G_i=g\} Y_i] \\
    &= &   \E[ \sum_{u \in \Delta^u_i(g)} \mo\{T_i=d, G_i=g, U_i=u\} Y_i(d,g,u)] \\
    &= &  \sum_{u \in \Delta^u_i(g)} \Pr(T_i=d, G_i=g, U_i=u) Y_i(d,g,u)
  \end{eqnarray*}
  where $\Delta_i^u(g)$ is the support $\{u: \Pr(U_i=u \mid G_i=g)
  >0\}.$ Therefore, the expectation of $\widehat{\ADE}_B$ is 
  \begin{eqnarray*}
    && \E[\widehat{\ADE}_B]   \\
    &  = & \frac{1}{N} \sum_{i=1}^N  \sum_{g \in \Delta^{g}_i} 
           \Pr (G_i=g) \biggl\{ \frac{\E[\mo\{T_i=1, G_i=g\} Y_i]}{\Pr(T_i=1,
           G_i=g)} -  \frac{\E[\mo\{T_i=0, G_i=g\} Y_i]}{\Pr(T_i=0,
           G_i=g)} \biggr\}  \\ 
    &= &  \frac{1}{N} \sum_{i=1}^N  \sum_{g \in \Delta^{g}_i} 
         \Pr (G_i=g) \Biggl\{ \\ 
    && \quad \sum_{u \in \Delta_i^u(g)} \biggl\{ \frac{\Pr(T_i=1, G_i=g, U_i=u) Y_i(1,g,u)}{\Pr(T_i=1,
       G_i=g)} -  \frac{\Pr(T_i=0, G_i=g, U_i=u) Y_i(0,g,u)}{\Pr(T_i=0,
       G_i=g)} \biggr\}  \Biggr\}\\ 
    &= &  \frac{1}{N} \sum_{i=1}^N  \sum_{g \in \Delta^{g}_i} 
         \Pr (G_i=g) \Biggl\{ \\ 
    && \quad \sum_{u \in \Delta_i^u(g)} \biggl\{ Y_i(1,g,u) \Pr(U_i=u \mid T_i=1,
       G_i=g)  - Y_i(0,g,u) \Pr(U_i=u \mid T_i=0, G_i=g)  \biggr\} \Biggr\}.
  \end{eqnarray*}
  Then, we have 
  \begin{eqnarray*}
    && \E[\hat{\ADE}_B]  - \ADE \\
    &= &\frac{1}{N} \sum_{i=1}^N \sum_{g \in \Delta^{g}_i} 
         \Pri (G_i=g) \biggl\{ \sum_{u \in \Delta_i^u} Y_i (1, g, u) 
         \{ \Pri(U_i=u \mid T_i=1, G_i=g) - \Pri(U_i=u \mid G_i=g)\}  \\
    && \quad -   \sum_{u \in \Delta_i^u} Y_i (0, g^\prime, u) \{ \Pr(U_i=u
       \mid T_i=0, G_i=g)  - \Pr(U_i=u \mid G_i=g) \}\biggr\}\\
    &= &\frac{1}{N} \sum_{i=1}^N \sum_{g \in \Delta^{g}_i} 
         \Pr (G_i=g) \biggl\{ \\ 
    && \quad \sum_{u \in \Delta_i^u(g)} \{Y_i (1, g, u) - Y_i(1, g, u^\prime)\} \{
       \Pr(U_i=u \mid T_i=1, G_i=g)  - \Pr(U_i=u \mid G_i=g)\}  \\
    &-& \quad \sum_{u \in \Delta_i^u(g)} \{Y_i (0, g^\prime, u) - Y_i(0,
        g^\prime, u^\prime)\} \{ \Pr(U_i=u
        \mid T_i=0, G_i=g)  - \Pr(U_i=u \mid G_i=g) \}\biggr\},
  \end{eqnarray*}
which completes the proof. \qed

\subsection{Proof of  Theorem~\ref{Parasen1}}
\label{app:sen1}
Using Theorem~\ref{biasASE}, under Assumption~\ref{linear}, 
\begin{eqnarray*}
  && \E[\hat{\ANSE}_B(\ag, \abg; d)]   - \ANSE(\ag, \abg; d) \\
  & = & \lambda \times \frac{1}{N} \sum_{i=1}^N \{\E[U_i=u \mid
        T_i=d, G_i=\ag] - \E[U_i=u \mid T_i=d, G_i=\abg]\}.
\end{eqnarray*}

From here, we focus on $\E[U_i \mid T_i=d, G_i=\ag]$. For notational
simplicity, we use $n_{G}(i)$ to denote the number of neighbors in the
network $\cG$ for individual $i$ and $n_U(i)$ is
similarly defined. Also, for individual $i$, let $\pi_{GU}(i)$ be the fraction of the
neighbors in $\cU$ who are neighbors in $\cG$ as well. Formally, 
$n_G(i) = |\cN^{\cG}_i|, n_U(i) = |\cN^{\cU}_i|$ and
$ \pi_{GU}(i) = |\cN^{(\cG,\cU)}_i| /|\cN^{\cU}_i|.$

First, we consider Bernoulli randomization with probability $p$. Under
this setting, 
\begin{eqnarray*}
  &&   \E[U_i \mid T_i=t, G_i=\ag] = \pi_{GU}(i) \times \ag + (1 - \pi_{GU}(i)) \times p. 
\end{eqnarray*}
Therefore, we have 
\begin{eqnarray*}
  && \E[\hat{\ANSE}_B(\ag, \abg; d)]   - \ANSE(\ag, \abg; d)  \\
  & = & \lambda \times \frac{1}{N} \sum_{i=1}^N \{\E[U_i=u \mid
        T_i=d, G_i=\ag] - \E[U_i=u \mid T_i=d, G_i=\abg]\}. \\ 
& = &   \lambda \times \frac{1}{N} \sum_{i=1}^N \pi_{GU}(i) \times (\ag - \abg)\\
& = &   \lambda \times \pi_{GU} \times (\ag - \abg).
\end{eqnarray*}
where the final equality follows from the definition of $\pi_{GU}$.

Next, we consider complete randomization with the number of treated
units $K$. Under this setting, 
\begin{eqnarray*}
&&   \E[U_i \mid T_i=d, G_i=\ag] \\
& = &
      \frac{n_{U}(i) \times \pi_{GU}(i) \times \ag }{n_U(i)} + (1 - \pi_{GU}(i))
      \frac{K- d - n_G(i) \times \ag}{N - 1 - n_{G}(i)}\\
& = & \pi_{GU}(i) \times \ag + (1 - \pi_{GU}(i))
      \frac{K- d - n_G(i) \times \ag}{N - 1- n_{G}(i)}\\
& = & \bigl\{ \pi_{GU}(i) - \frac{n_G(i)}{N-1-n_G(i)}
      (1-\pi_{GU}(i)) \bigr\} \ag + \frac{K-d}{N-1-n_{G}(i)}(1 - \pi_{GU}(i))
\end{eqnarray*}
When $N$ is much larger than $n_G(i)$,  \ $ n_G(i) /(N - 1 - n_{G}(i)) \approx
0 \label{eq:approx}$. Then,  we have 
\begin{eqnarray*}
&&   \E[U_i \mid T_i=d, G_i=\ag]  \approx  \pi_{GU}(i) \ag +
   \frac{K-d}{N-1-n_{G}(i)}(1 - \pi_{GU}(i)), \\
&&   \E[U_i \mid T_i=d, G_i=\ag]  - \E[U_i=u \mid T_i=d, G_i=\abg]
   \approx \pi_{GU}(i) (\ag - \abg).
\end{eqnarray*}

Therefore, when $N$ is much larger than $n_G(i)$ for all $i$, we get 
the simplified bias formula.
\begin{eqnarray*}
  && \E[\hat{\ANSE}_B(\ag, \abg; d)]   - \ANSE(\ag, \abg; d)  \\
  & = & \lambda \times \frac{1}{N} \sum_{i=1}^N \{\E[U_i=u \mid
        T_i=d, G_i=\ag] - \E[U_i=u \mid T_i=d, G_i=\abg]\}. \\ 
& \approx &   \lambda \times \frac{1}{N} \sum_{i=1}^N \pi_{GU}(i) \times (\ag - \abg)\\
& = &   \lambda \times \pi_{GU} \times (\ag - \abg).
\end{eqnarray*}

Finally, we consider a situation when $N$ is not large enough to have the
aforementioned approximation. Suppose $N \approx (C+1) n_G(i) + 1$ for all
$i$.  Then, 
\begin{eqnarray*}
  &&   \E[U_i \mid T_i=d, G_i=\ag]  \approx  \bigl\{ \frac{C+1}{C} \pi_{GU}(i) - \frac{1}{C} \bigr\} \times \ag  +
   \frac{K-d}{N-1-n_{G}(i)}(1 - \pi_{GU}(i)),  \\
  &&   \E[U_i \mid T_i=d, G_i=\ag]  - \E[U_i=u \mid T_i=d, G_i=\abg] \approx
     \bigl\{ \frac{C+1}{C} \pi_{GU}(i) - \frac{1}{C}  \bigr\} (\ag - \abg). 
\end{eqnarray*}

Therefore, the bias can be written as,
\begin{eqnarray}
  && \E[\hat{\ANSE}_B(\ag, \abg; d)]   - \ANSE(\ag, \abg; d)
     \notag \\
  & = & \lambda \times \frac{1}{N} \sum_{i=1}^N \{\E[U_i=u \mid
        T_i=d, G_i=\ag] - \E[U_i=u \mid T_i=d, G_i=\abg]\}. \notag \\ 
& \approx &   \lambda \times \bigl\{ \frac{C+1}{C} \times \frac{1}{N} \sum_{i=1}^N
      \pi_{GU}(i) - \frac{1}{C} \bigr\} \times (\ag - \abg). \notag \\
& = &   \lambda \times \bigl\{ \frac{C+1}{C} \pi_{GU} - \frac{1}{C}
      \bigr\}\times (\ag - \abg). \label{eq:sen1}
\end{eqnarray}
\qed
\subsection{Proof of Theorem~\ref{sen2}}
\label{app:sen2}
First, we set the following notations. We define the support
$\Delta^u_s$ to be the support $\Delta_i^u$ for all $i$ with $S_i=s$.
We drop subscript $s$ whenever it is obvious from contexts. For $\bg
\in \gs$,
\begin{eqnarray*}
  r_{\bg} (u)  & \equiv & \frac{1}{N} \sum_{i: S_i=s} Y_i(d, \bg, u) \\
  v_{\ag} (\bg) & \equiv & \frac{\sum_{u \in \Delta_s^u} \{ r_{\bg} (u) - \minb_u r_{\bg} (u) \}
                       \ \Pr(U_i=u \mid T_i=d, G_i=\ag)}{\maxb_u r_{\bg} (u)  -
                       \minb_u r_{\bg} (u) } \\
  v_{\abg} (\bg) & \equiv & \frac{\sum_{u \in \Delta_s^u} \{ r_{\bg} (u) - \minb_u r_{\bg} (u) \}
                      \Pr(U_i=u \mid T_i=d, G_i=\abg)}{\maxb_u r_{\bg} (u)  -
                           \minb_u r_{\bg} (u) } \\
  \Gamma (\bg)
               & \equiv & \frac{v_{\ag} (\bg) }{v_{\abg}(\bg)} = 
                          \frac{\sum_{u \in \Delta_s^u} \{ r_{\bg} (u) - \minb_u r_{\bg} (u) \}
                          \Pr(U_i=u \mid T_i=d, G_i=\ag)}{\sum_{u \in \Delta_s^u}
                          \{ r_{\bg} (u) - \minb_u r_{\bg} (u) \} \Pr(U_i=u \mid T_i=d, G_i=\abg)} \\
  MR^{obs}(\ag, \abg ; s) & \equiv &  
                                       \frac{\sum_{u \in \Delta_s^u} r_{\ag}
                                       (u) \Pr(U_i=u \mid T_i=d, G_i=\ag)}{\sum_{u \in \Delta_s^u}
                                       r_{\abg} (u) \Pr(U_i=u \mid T_i=d, G_i=\abg)}\\
  MR^{true}_{\bg}(\ag, \abg; s) & \equiv & \frac{\sum_{u \in
                                             \Delta_s^u} r_{\ag} (u)
                                             \Pr(U_i=u \mid T_i=d, G_i=\bg)}
                                             {\sum_{u \in \Delta_s^u} r_{\abg} (u) \Pr(U_i=u \mid T_i=d, G_i=\bg)}
\end{eqnarray*}
where $0 \leq v_{\ag} (\bg), v_{\abg} (\bg)\leq  1$ because of non-negative
outcomes. 

\begin{lemma}
  \label{senProof}
  For $(\ag, \abg)$,
\begin{eqnarray*}
  \frac{MR^{obs}(\ag, \abg;s)}{MR^{true}_{\abg}(\ag, \abg; s)}
  \ \leq \   B \ \ \ 
  & & \ \ \ \frac{MR^{obs} (\ag, \abg; s)}{MR^{true}_{\ag}(\ag, \abg; s)}
      \ \leq \   B,\\
  \frac{MR^{obs}(\abg, \ag;s)}{MR^{true}_{\abg}(\abg, \ag; s)}
  \ \leq \   B \ \ \ 
  && \ \ \  \frac{MR^{obs} (\abg, \ag; s)}{MR^{true}_{\ag}(\abg, \ag; s)}
   \ \leq \   B.
\end{eqnarray*}
\end{lemma}

\paragraph{Proof}
This proof closely follows \cite{ding2016sensitivity}. The key difference is that we study bias due to an unmeasured
relevant network in the presence of interference in multiple networks in
contrary to bias due to an unmeasured confounder in observational
studies without interference \citep{ding2016sensitivity}. 

For $\bg \in \gs$ and $s$,
\begin{eqnarray*}
  \Gamma (\bg) & = &  \frac{\sum_{u \in \Delta_s^u} \{ r_{\bg} (u) - \minb_u r_{\bg} (u) \}
                     \Pr(U_i=u \mid T_i=d, G_i=\ag)}{\sum_{u \in \Delta_s^u}
                     \{ r_{\bg} (u) - \minb_u r_{\bg} (u) \} \Pr(U_i=u \mid T_i=d, G_i=\abg)} \\
               & = & \frac{\sum_{u \in \Delta_s^u} \{ r_{\bg} (u) - \minb_u r_{\bg} (u) \}
                     \frac{ \Pr(U_i=u \mid T_i=d, G_i=\ag)}{ \Pr(U_i=u \mid T_i=d, G_i=\abg)}  \Pr(U_i=u \mid T_i=d, G_i=\abg)}
                     {\sum_{u \in \Delta_s^u} \{ r_{\bg} (u) - \minb_u r_{\bg} (u) \} \
                      \Pr(U_i=u \mid T_i=d, G_i=\abg)} \\
               &  \leq & \RR_{GU}
\end{eqnarray*}
Also, for $\bg \in \gs$ and $s$,
\begin{eqnarray*}
  \frac{1}{\Gamma (\bg)} & = &  \frac{\sum_{u \in \Delta_s^u} \{ r_{\bg} (u) - \minb_u r_{\bg} (u) \}
                               \Pr(U_i=u \mid T_i=d, G_i=\abg)}{\sum_{u \in \Delta_s^u}
                               \{ r_{\bg} (u) - \minb_u r_{\bg} (u) \} \Pr(U_i=u \mid T_i=d, G_i=\ag)} \\
                         & = & \frac{\sum_{u \in \Delta_s^u} \{ r_{\bg} (u) - \minb_u r_{\bg} (u) \}
                               \frac{ \Pr(U_i=u \mid T_i=d, G_i=\abg)}{
                               \Pr(U_i=u \mid T_i=d, G_i=\ag)} 
                               \Pr(U_i=u \mid T_i=d, G_i=\ag)}
                               {\sum_{u \in \Delta_s^u} \{ r_{\bg} (u) - \minb_u r_{\bg} (u) \} \
                               \Pr(U_i=u \mid T_i=d, G_i=\ag)} \\
                         &  \leq & \RR_{GU}.
\end{eqnarray*}
Then, we have 
\begin{eqnarray*}
  && \frac{MR^{obs}(\ag, \abg; s)}{MR^{true}_{\abg}(\ag, \abg;
  s)} \\
  & = & \frac{\sum_{u \in \Delta_s^u} r_{\ag} (u) \Pr(U_i=u \mid T_i=d,
        G_i=\ag)}{\sum_{u \in \Delta_s^u} r_{\abg} (u) \Pr(U_i=u \mid T_i=d, G_i=\abg)}
        \times \frac{\sum_{u \in \Delta_s^u} r_{\abg} (u) \Pr(U_i=u
        \mid T_i=d, G_i=\abg)}
        {\sum_{u \in \Delta_s^u} r_{\ag} (U) \Pr(U_i=u \mid T_i=d, G_i=\abg)}\\
  & = &  \frac{\sum_{u \in \Delta_s^u} r_{\ag} (u) \Pr(U_i=u \mid T_i=d,
        G_i=\ag)}{\sum_{u \in \Delta_s^u} r_{\ag} (u) \Pr(U_i=u \mid T_i=d, G_i=\abg)}\\
  & = & \frac{\{ \maxb_u r_{\ag} (u)  - \minb_u r_{\ag} (u)\} v_{\ag} (\ag) + \minb_u r_{\ag} (u)}{\{ \maxb_u r_{\ag} (u)  - \minb_u r_{\ag} (u)\} \frac{v_{\ag} (\ag)}{\Gamma(\ag)} + \minb_u r_{\ag} (u)}
\end{eqnarray*}
From Lemma A.1 in \cite{ding2016sensitivity},
when $\Gamma(\ag) > 1$,
$  \frac{\MR^{obs}(\ag, \abg;s)}{\MR^{true}_{\abg}(\ag, \abg;s)}$
is increasing in $v_{\ag}(\ag)$. Therefore, it takes the maximum value when
$v_{\ag}(\ag) = 1$. 
\begin{eqnarray*}
  \frac{\MR^{obs}(\ag, \abg;s)}{\MR^{true}_{\abg}(\ag, \abg;s)}
  & \leq & \frac{ \Gamma (\ag) \times \MR_{UY}(\ag,s)}{\Gamma (\ag) + \MR_{UY}(\ag,s) - 1}\\
  & \leq & \frac{ \RR_{GU} \times \MR_{UY}}{ \RR_{GU} + \MR_{UY} - 1}
\end{eqnarray*}
where the second inequality comes from Lemma A.2 in
\cite{ding2016sensitivity} and $\Gamma (\ag) \leq \RR_{GU}, \MR_{UY} =
\maxb_{g,s} \MR_{UY}(g,s)$. 

From Lemma A.1 in \cite{ding2016sensitivity},
when $\Gamma(\ag) \leq 1$,
$  \frac{\MR^{obs}(\ag, \abg;s)}{\MR^{true}_{\abg}(\ag, \abg;s)}$
is non-increasing in $v_{\ag}(\ag)$. Therefore, it takes the maximum value
at $v_{\ag}(\ag) = 0$. 
$$ \frac{\MR^{obs}(\ag, \abg;s)}{\MR^{true}_{\abg}(\ag, \abg;s)}
\leq  1 \leq  \frac{ \RR_{GU} \times \MR_{UY}}{ \RR_{GU} + \MR_{UY} - 1} $$
where the second inequality comes from Lemma A.2 in
\cite{ding2016sensitivity} and $ \RR_{GU} \geq 1 , \MR_{UY} \geq 1$.

Hence, we obtain the desired result. 
\begin{equation*}
  \frac{\MR^{obs}(\ag, \abg;s)}{\MR^{true}_{\abg}(\ag, \abg;s)}
  \leq  \frac{ \RR_{GU} \times \MR_{UY}}{ \RR_{GU} + \MR_{UY} - 1}.
\end{equation*}
Similar derivations apply to the other three inequalities. \qed

\paragraph{Proof of the theorem.}
For notational simplicity, we use the following representation.
\begin{eqnarray*}
  m(d,g; s)  & \equiv & \frac{1}{N} \sum_{i: S_i=s} \E[Y_i \mid T_i=d, G_i =  g]\\
                   & = & \frac{1}{N} \sum_{i: S_i=s} \frac{\sum_{u \in
                         \Delta_s^u} \Pr(T_i=d, G_i=g, U_i=u) Y_i(d,
                         g, u)}{\Pr(T_i=d, G_i=g)}\\
                   & = & \frac{1}{N} \sum_{i: S_i=s} \sum_{u \in \Delta_s^u} \Pr(U_i=u \mid T_i=d, G_i=g) Y_i(d, g, u)\\
                   & = & \sum_{u \in \Delta_s^u} \biggl\{ \frac{1}{N} \sum_{i:
                         S_i=s} Y_i(d, g, u) \biggr\} \Pr(U_i=u \mid
                         T_i=d, G_i=g)\\
& = & \sum_{u \in \Delta_s^u} r_g(u) \Pr(U_i=u \mid T_i=d, G_i=g).
\end{eqnarray*}

We want to show that,  for $\ag, \abg$,
\begin{eqnarray*}
  \frac{m(d,\ag;s)}{B} - B \times m(d, \abg; s) \ \leq \
  \frac{1}{N} \sum_{i: S_i=s} \NSE_i(\ag, \abg; d) \
  \leq   B \times m(d,\ag;s) - \frac{m(d, \abg; s)}{B}.
\end{eqnarray*}
Because this implies the desired result.
\begin{eqnarray*}
\hspace{-0.5in} &  &  \frac{m(d,\ag;s)}{B} - B \times m(d, \abg; s) \ \leq \
                     \frac{1}{N} \sum_{i: S_i=s} \NSE_i(\ag, \abg; d) \
                     \leq   B \times m(d,\ag;s) - \frac{m(d, \abg; s)}{B} \\
  \hspace{-0.5in} & \Leftrightarrow & 
                                      \left\{
                                      \begin{array}{ll}
                                        \sum_{s \in \cS} \biggl\{
                                        \frac{m(d,g;s)}{B} - B \times m(d, \abg; s) \biggr\} \ \leq \
                                        \sum_{s \in \cS}\biggl\{\frac{1}{N} \sum_{i: S_i=s} \NSE_i(\ag, \abg; d) \biggr\} \\
                                        \sum_{s \in
                                        \cS}\biggl\{\frac{1}{N}
                                        \sum_{i: S_i=s} \NSE_i(\ag,
                                        \abg; d) \biggr\} \ \leq \
                                        \sum_{s \in \cS} \biggl\{ B
                                        \times m(d,\ag;s) -
                                        \frac{m(d, \abg; s)}{B}
                                        \biggr\}
                                      \end{array}
  \right.\\
  \hspace{-0.5in} & \Leftrightarrow &  \frac{\E[\hat{m}(d,\ag)]}{B} - B
                                      \times \E[\hat{m}(d, \abg)] \ \leq \
                                      \ANSE(\ag, \abg; d) \
                                      \leq   B \times \E[\hat{m}(d,\ag)] - \frac{\E[\hat{m}(d, \abg)]}{B}. 
\end{eqnarray*}

First, using Lemma~\ref{senProof}, 
\begin{eqnarray*}
  &&  \frac{m(d, \ag; s)}{ \sum_{u \in \Delta_s^u} r_{\ag}(u) \Pr(U_i=u \mid T_i=d, G_i=\abg)}\\
  & = & \frac{  \sum_{u \in \Delta_s^u} r_{\ag}(u) \Pr(U_i=u \mid T_i=d,
        G_i=\ag)} {\sum_{u \in \Delta_s^u} r_{\ag}(u) \Pr(U_i=u \mid T_i=d, G_i=\abg)}\\
  & = & \frac{ \sum_{u \in \Delta_s^u} r_{\ag}(u) \Pr(U_i=u \mid T_i=d,
        G_i=\ag)}
        {\sum_{u \in \Delta_s^u} r_{\abg} (u) \Pr(U_i=u \mid T_i=d,
        G_i=\abg)} \ 
        \times \frac{\sum_{u \in \Delta_s^u} r_{\abg}(u) \Pr(U_i=u \mid T_i=d, G_i=\abg)}{\sum_{u \in \Delta_s^u} r_{\ag}(u)
        \Pr(U_i=u \mid T_i=d, G_i=\abg)}\\
  & = &  \frac{MR^{obs}(\ag, \abg; s)}{MR^{true}_{\abg}(\ag, \abg; s)}  \leq   B
\end{eqnarray*}
where the final equality follows from the lemma. 
Therefore, 
\begin{equation}
  \frac{m(d, \ag; s)}{B} \ \leq \  \sum_{u \in \Delta_s^u} r_{\ag}(u) \Pr(U_i=u \mid T_i=d, G_i=\abg).  \label{eq:sen21}
\end{equation}

Also, since $B \geq 1$,  
\begin{equation}
  \sum_{u \in \Delta_s^u} r_{\abg} (U) \Pr(U_i=u \mid T_i=d, G_i=\abg) \
  = m(d, \abg; s) \ \leq \ B \times m(d, \abg; s). \label{eq:sen22}
\end{equation}
Finally, taking equations~\eqref{eq:sen21} and \eqref{eq:sen22}
together, 
\begin{eqnarray*}
  && \frac{m(d, \ag; s)}{B} - B \times m(d, \abg; s) \\
  & \leq  & \sum_{u \in \Delta_s^u} r_{\ag}(u) \Pr(U_i=u \mid T_i=d,
           G_i = \abg) - \sum_{u \in \Delta_s^u} r_{\abg}(u) \Pr(U_i=u \mid T_i=d,
           G_i = \abg)\\
& =  & \sum_{u \in \Delta_s^u} \{r_{\ag}(u) - r_{\abg}(u)\} \Pr(U_i=u \mid T_i=d,
          G_i =  \abg)\\
& =  & \frac{1}{N} \sum_{i: S_i=s} \sum_{u \in \Delta_s^u} \{Y_i(d,
       \ag,u) - Y_i(d, \abg, u)\} \Pr(U_i=u \mid T_i=d, G_i = \abg) \\
& =  & \frac{1}{N} \sum_{i: S_i=s} \NSE_i(\ag, \abg; d).
\end{eqnarray*}

Similarly, we want to prove
\begin{eqnarray*}
  && \frac{1}{N} \sum_{i: S_i=s} \NSE_i(\ag, \abg; d) \leq B \times m(d, \ag; s) - \frac{m(d, \abg; s)}{B}.
\end{eqnarray*}

First, 
since $B \geq 1$,  
\begin{equation}
  \frac{m(d, \abg; s)}{B} \ \leq \  m(d, \abg; s) \ = \ \sum_{u \in \Delta_s^u} r_{\abg}(u) \Pr(U_i=u \mid T_i=d, G_i=\abg).  \label{eq:sen31}
\end{equation}
Then, using Lemma~\ref{senProof}, 
\begin{eqnarray*}
  &&  \frac{\sum_{u \in \Delta_s^u} r_{\ag}(u) \Pr(U_i=u \mid T_i=d,G_i=\abg)}{m(d, \ag; s)}\\
  & = & \frac{ \sum_{u \in \Delta_s^u} r_{\ag}(u) \Pr(U_i=u \mid
        T_i=d, G_i=\abg)}{\sum_{u \in \Delta_s^u} r_{\ag}(u) \Pr(U_i=u \mid T_i=d,G_i=\ag)}\\
  & = & \hspace{-0.1in}\frac{\sum_{u \in \Delta_s^u} r_{\ag}(u) \Pr(U_i=u \mid T_i=d,
           G_i=\abg)} {\sum_{u \in \Delta_s^u} r_{\abg} (u) \Pr(U_i=u \mid T_i=d,
           G_i=\abg)} \times \frac{\sum_{u \in \Delta_s^u} r_{\abg}(u) \Pr(U_i=u \mid T_i=d,
           G_i=\abg)}{ \sum_{u \in \Delta_s^u} r_{\ag}(U) \Pr(U_i=u \mid T_i=d,
           G_i=\ag)}\\
  & = &  \frac{MR^{obs} (\abg, \ag; s)}{MR^{true}_{\abg}(\abg, \ag; s)} \leq   B.
\end{eqnarray*} 
Therefore, we have 
\begin{equation}
  \sum_{u \in \Delta_s^u} r_{\ag} (U) \Pr(U_i=u \mid T_i=d, G_i=\abg)
  \ \leq \ B \times m(d, \ag; s). \label{eq:sen32}
\end{equation}


Finally, taking equations~\eqref{eq:sen31} and \eqref{eq:sen32}
together, 
\begin{eqnarray*}
  && B \times m(d, \ag; s) - \frac{m(d, \abg; s)}{B} \\
  & \geq  & \sum_{u \in \Delta_s^u} r_{\ag}(u) \Pr(U_i=u \mid T_i=d,
           G_i = \abg) - \sum_{u \in \Delta_s^u} r_{\abg}(u) \Pr(U_i=u \mid T_i=d,
           G_i = \abg)\\
& =  & \sum_{u \in \Delta_s^u} \{r_{\ag}(u) - r_{\abg}(u)\} \Pr(U_i=u \mid T_i=d,
          G_i = \abg)\\
& =  & \frac{1}{N} \sum_{i: S_i=s} \sum_{u \in \Delta_s^u} \{Y_i(d,
       \ag,u) - Y_i(d, \abg, u)\} \Pr(U_i=u \mid T_i=d, G_i = \abg) \\
& =  & \frac{1}{N} \sum_{i: S_i=s} \NSE_i(\ag, \abg; d).
\end{eqnarray*}
Hence we have
\begin{eqnarray*}
  \frac{m(d,\ag;s)}{B} - B \times m(d, \abg; s) \ \leq \
  \frac{1}{N} \sum_{i: S_i=s} \NSE_i(\ag, \abg; d) \
  \leq   B \times m(d,\ag;s) - \frac{m(d, \abg; s)}{B},
\end{eqnarray*}
which completes the proof. \qed

\end{document}